\documentclass[twocolumn,showpacs,preprintnumbers,amsmath,amssymb]{revtex4}

\usepackage{graphicx}% Include figure files
\usepackage{dcolumn}% Align table columns on decimal point
\usepackage{bm}% bold math

\begin{document}

%\preprint{APS/123-QED}

\title{Transport on Adaptive Random Lattices}% Force line breaks with \\

\author{Jelle \surname{Ritzerveld}}
 	\email{ritzerveld@strw.leidenuniv.nl}
%  	\homepage{http://www.strw.leidenuniv.nl/~ritzerveld}
\author{Vincent \surname{Icke}}
	\affiliation{Sterrewacht Leiden, P.O. Box 9513, 2300 RA Leiden, The Netherlands}%

\date{\today}

\begin{abstract}

In this paper, we present a new method for the solution of those linear transport processes that may be described by a Master Equation, such as electron, neutron and photon transport, and more exotic variants thereof. We base our algorithm on a Markov process on a Voronoi-Delaunay grid, a nonperiodic lattice which is derived from a random point process that is chosen to optimally represent certain properties of the medium through which the transport occurs. Our grid is locally translation and rotation invariant in the mean. We illustrate our approach by means of a particular example, in which the expectation value of the length of a grid line corresponds to the local mean free path. In this example, the lattice is a direct representation of the `free path space' of the medium. Subsequently, transport is defined as simply moving particles from one node to the next, interactions taking place at each point. We derive the statistical properties of such lattices, describe the limiting behavior, and show how interactions are incorporated as global coefficients. Two elementary linear transport problems are discussed: that of free ballistic transport, and the transport of particles through a scattering medium. We also mention a combination of these two. We discuss the efficiency of our method, showing that it is much faster than most other methods, because the operation count does \emph{not} scale with the number of sources. We test our method by focusing on the transport of ionizing radiation through a static medium, and show that the computed results for the classical test case of an ionization front expanding in a homogeneous medium agree perfectly with the analytic solution. We finish by illustrating the efficiency and flexibility of our method with the results of a simulation of the reionization of the large scale structure of the Universe.

\end{abstract}


% 05.10. Gg = Stochastic models in statistical physics and nonlinear dynamics
% 51.10.+y   = Transport processes in gases
% 95.30.Jx   = Radiative transfer in astrophysics

\maketitle

\section{Introduction}

\subsection{Grids and errors}

The numerical description of physical systems that obey a differential expression requires the use of a differencing technique on some sort of grid, mesh or lattice (except in rare cases where computational symbolic algebra can be applied). Accordingly, the question is not if errors are made, but rather what kind, and of what severity. In the past, computational lattices were almost always regular and rectangular, the Cartesian grid being the archetypal example. Analytic estimates of the corresponding errors are routinely made for ordinary differential equations on such grids (e.g. \cite{NumRecipes}); in the case of partial differential equations, especially those of hydrodynamics, the error analysis may even be quite elegant (see e.g. \cite{RichtmyerMorton}, \cite{VanLeerThesis}, \cite{IckeBubbles}).

In almost all of these cases, the numerical method may be chosen in such a way that the errors become small with increasing grid resolution. In practice this is often less than helpful, due to the steep increase of computational effort with decreasing mesh size. Moreover, there is one type of error that does not automatically vanish, namely those effects that are due to the geometry of the grid. Any intrinsic regularity (such as is introduced by using a Cartesian grid) will, by Noether's Theorem, produce spurious conservation laws. These are sometimes innocuous, but may be rather vicious under some circumstances. One of the simplest examples is that of the Short Characteristics method \cite{Kunasz} in radiation transport, in which the numerical diffusion along the axes is negligible compared to that along the diagonals, resulting in spiky features along the axes when modeling isotropic outflow.

It is this deficiency in particular that we wish to address here. The solution seems to be obvious: pick an irregular grid that is locally isotropic in the mean. The question then immediately arises, how to construct such a grid, and especially how to design its properties (and the corresponding algorithm that describes the physical process). 

In this paper, we present the following procedure for the construction of a computational grid. First, we represent the transport medium by a point distribution. Second, these points serve as nodes in a Delaunay triangulation \cite{Okabe}, a much used tiling of space described in Section III. Third, we specify the physical operators that act along the Delaunay lines that connect the points (in our specific example, these operators constitute a Markov transport process). The whole procedure is carried out in such a way that the point distribution (often called {\it point process\/} in the mathematics literature) optimizes the solution. By `representing' we might mean a Poisson process with $N$ points with an intensity $n_\mathrm{p}(\vec x)$ in 3-space, such that for $N\rightarrow \infty$ the function $n_\mathrm{p}$ is proportional to the matter density $n$. A more general case, which we will fully exploit below, is a Poisson process with $n_\mathrm{p}=f(n)$, in which the function $f$ is designed to optimize certain transport properties of the algorithm.

We believe that this approach is extremely general and flexible, and may be used profitably in statistical physics, hydrodynamics and even quantum mechanics (cf.\ \cite{Christ}). In this paper, however, we focus only on \emph{linear} transport problems, of particles moving through a background medium. Thus, we ignore any nonlinear particle-particle interaction terms. For definiteness we choose to illustrate it here by means of one specific example, namely the transport of radiation through a medium that is optically active (scattering, absorption, ionization). In that case, the `transport medium' consists of gas and dust (in other cases the Delaunay lines might represent, say, communication channels, or gluons connecting quarks).

\subsection{Transport processes}

The Master Equation (ME) is the mainstay of stochastic processes \cite{Kampen}, and can be used, in its most general form, to describe the redistribution of probability in some abstract space. It relates the rate of change of some density, or distribution function, to one or more gain and loss terms which describe interactions, in the most general sense. A very successful version of the ME is the versatile Boltzmann Equation (BE), widely used in transport theory  \cite{Cercignani}. In abstract form, it reads:

\begin{equation}
  \label{EqAbstractBoltzmann}
  \mathbf{D}f=\mathbf{C}f,
\end{equation}
in which \(\mathbf{D}\) is a drift operator and \(\mathbf{C}\) a collision operator, and \(f\) is a probability density, well-defined on a certain region of this abstract space. 

One can project the operators in Eq.(\ref{EqAbstractBoltzmann}) onto the phase space \(\Gamma\), to obtain one particular form of the BE, that can be used to describe the flow of particles through an active medium with which they might interact. In the absence of external forces,
\begin{equation}
  \label{EqLinearBoltzmann}
  \left[\frac{\partial}{\partial t}+\vec{v}\cdot\vec{\nabla}\right]f(\vec{\mu})=\left. \frac{\partial f(\vec{\mu})}{\partial t}\right|_\mathrm{coll},
\end{equation}
in which \(f(\vec{\mu})=f(\vec{x},\vec{v},E,t)\) is the probability density for particles at position \(\vec{x}\), with velocity \(\vec{v}\), with an energy \(E\), at time t. The collision terms on the right hand side of Eq.(\ref{EqLinearBoltzmann}) can be written down more explicitly in terms of gain and loss terms, once the types of interactions are known. In the case of scattering one would obtain:

\begin{eqnarray}
  \label{EqGainLoss}
  \left. \frac{\partial f(\vec{x},\vec{v},E,t)}{\partial t}\right|_\mathrm{coll}&=&\int_{4\pi}\mathrm{d}\vec{v}'\Sigma(\vec{v}\cdot\vec{v}', E)f(\vec{x},\vec{v}',E,t)\nonumber\\
  &-&f(\vec{x},\vec{v},E,t)\int_{4\pi}\mathrm{d}\vec{v}'\Sigma(\vec{v}\cdot\vec{v}', E),
\end{eqnarray}
in which \(\Sigma(\vec{v}\cdot\vec{v}', E)\) is the macroscopic differential scattering cross section, equivalent to the reciprocal of the local mean free path. In general, additional particle-particle and particle-medium interactions can be added to the collision term, each interaction having its own cross section, contributing to the total redistribution kernel.

Originally used to describe the behaviour of classical gases, the generalized BE has shown its use in a wide variety of transport processes, from that of the transport of neutrons in nuclear reactors, to that of photons in superfluids, of radiation through stellar atmospheres, to even the behavioural patterns of humans in a large crowd. As such, linear and non-linear representations of Eq.(\ref{EqAbstractBoltzmann}) have been analyzed rigorously \cite{Bellomo}, but in almost all cases it is nigh to impossible to find closed analytic solutions, even more so if one wants to find time-dependent solutions. It is therefore of the utmost scientific importance to develop reliable, fast and flexible numerical methods, which are able to deal with the complexities of transport theoretic problems.

We stress again that, in this paper, we present our method only as a tool for solving linear transport problems, such as the transport of electrons, neutrons and photons through a background medium. Extensions to the nonlinear regime are possible, but we will only hint at that in Sect.\ \ref{subsec:CorrelationFunction}. 

\subsection{Numerical methods}

Extant numerical transport methods come in a wide variety of forms, depending on which type of transport process Eq.(\ref{EqAbstractBoltzmann}) represents. In general, the computational techniques can be split into two categories: deterministic and stochastic methods. Because of the advent of ever more powerful computers, stochastic methods have been developed which mimic the behaviour of the flow itself. Monte Carlo methods based on random Markov processes are very popular, because they are conceptually simple and they are very easy to adapt to parallel processing, considering they are local. Most notably, the Direct Simulation Monte Carlo (DSMC) \cite{Bird}, and variants thereof, have been widely used to study the dynamics of rarefied gases where the Knudsen number (\(\lambda/L\), a dimensionless parameter relating the local mean free path to the dimension of the volume) is high.

\subsubsection{Structured lattices}

In almost all variants of these stochastic methods, one constructs a grid, or a lattice, on which one does not only have to discretise the distribution function (and the differential operators), but also the medium properties, and thus the interaction coefficients. The transport can henceforth be solved by, for example, tracing rays (radiation transport) or transporting fluxes (hydro-solvers) from one grid cell to the next. Although a regular mesh might seem the most obvious choice, it is known to cause several problems, inherent to its structure. If the medium is highly inhomogeneous, the global resolution of a stiff rectangular mesh has to be high enough to be able to sample the highest Fourier component of the density spectrum. This, of course, results in a high redundancy of grid cells in fairly homogeneous regions. This problem has been partly solved by the introduction of Adaptive Mesh Refinement (AMR) \cite{AMR1, AMR2}, in which the grid refinement follows some criterion, such as the gradient in the density. Even refined grids, though, suffer from another problem all regular grids have: they are known to break physical symmetries. Because the cells are congruent, albeit not of the same size, the rotation group of the lattice is not isomorphic to \(SO(d)\), but to some discrete subgroup thereof. Also, the fixed widths of the cells impose a non-physical constraint, and are known to break translational symmetry. Several different communities have described these shortcomings in different forms. DSMC methods require the mesh size to be much smaller than the local mean free path, in order to accurately capture the flow features. When the density of the gas locally becomes very high, the computational cost will increase tremendously, imposing a difficulty for extending DSMC methods to the continuum limit, where the Knudsen number \(\lambda/L\rightarrow 0\). The Lattice Boltzmann (LB) community have shown \cite{FHP} that the discrete rotation group of a rectangular lattice does not have enough symmetry to obtain Navi\'{e}r Stokes-like equations in the limit, and they had to resort to hexagonal lattices in 2D and multi-speed models in 3D, in the absence of Platonic solids that at the same time have a large enough symmetry group \emph{and} tessellate space \cite{Wolfe}. Moreover, the regular lattices are known to break Galilean symmetry \cite{Succi}. And, the fixed cell-widths impose constraint on the Reynolds number the method can resolve. In the lattice gauge community, it has been known for quite some time that regular (Wilson) lattices break Poincar\'{e} symmetry \cite{Christ}. Supersymmetry closes on the Poincar\'{e} group by necessity, and therefore has difficulty being defined on regular lattices \cite{Kaku}. Moreover, because the lattices are invariant under translations of one or more cell-widths, or rotations of \(\pi/2\) (or \(\pi/3\) when one uses triangular grids) with respect to one of the axes, spurious invariants are introduced. 

In summary, the choice of a regular grid which has nothing to do with the underlying physical problem results in the introduction of unphysical conserved quantities, and the breaking of several very physical symmetries.

\subsubsection{Random lattices}

Various different communities have independently found an answer to this mesh-related problem. Dispense with the regular grids altogether, and introduce \emph{random} lattices, which will be described in more detail later on in this paper. In those fields of physics, where symmetries are most important, the lattices were used first. General relativity was discretised onto a simplicial lattice \cite{Regge}, even resulting in quantum gravity theories, and lattice gauge theories were defined on similar random lattices \cite{Christ}. Their use was hinted at in cellular automaton fluids  \cite{Wolfram}, but was overlooked as a possibility to solve the dichotomy between symmetric and space-filling lattices in LB solvers (for a review on unstructured grids in LB methods, cf.\ \cite{Ubertini}). Voronoi lattices were also recognized to be of use in the field of Dissipative Particle Dynamics \cite{Espanol, Flekkoy}. These random lattices still have a problem, though. In most cases, a Poisson point process lies at their basis, a result of which is that the average point-to-point distance is homogeneously of the order of \(n_\mathrm{p}^{-1/d}\), in which \(n_\mathrm{p}\) is the density of points, and \(d\) is the dimension. For an inhomogeneous medium distribution this length scale does not have an immediate correspondence to the length scales of the physical problem, and, as said, impose constraints on the physical parameters (i.e. the Reynolds number for LB methods and the Knudsen for DSMC methods) to be resolved.

\subsection{Outline}

In this paper, we describe a new linear transport method which dispenses with the unphysical regular meshes, and uses as a basis random lattices that are locally isotropic in the mean. Moreover, the point distribution which defines the lattice represents the medium distribution, from which it follows that the length scales introduced (e.g. the mean Delaunay edge length) are not irrelevant, or just a step in a refinement sequence, but have true physical meaning. The method we will describe can be used more general to solve almost all MEs of type Eq.(\ref{EqAbstractBoltzmann}), and can be used as such, but in this paper we will specifically focus on linear transport processes of particles being transported through a (possibly dynamically evolving) background medium, with which they interact. Nonlinear particle-particle interactions are ignored, except through feedback from the background medium itself. Thus, the method as described in this paper can be used to model transport processes such as electron, neutron or photon transport. We will end this paper by focusing on one particular example of such particle transport, namely that of photon transport, and we will compare how an implementation of our method compares to analytical solutions. Hereafter, we show its flexibility in a simulation of the Epoch of Reionization in the early Universe, in which the first generation of stars cause a phase transition of hydrogen gas in the Universe, namely from neutral to almost fully ionized.

\section{Stochastic methods}
A robust approach to numerically solving linear transport problems is the use of Monte Carlo methods, in which the solution for a macroscopic system is obtained by randomly sampling microscopic interactions. Because our new method bears many similarities to standard stochastic methods, we proceed by first pointing out the essentials, after which we will describe our new method.

\subsection{Monte Carlo transport methods}

In very general terms, stochastic transport methods are set up to solve Eq.(\ref{EqAbstractBoltzmann}) as follows. The BE is split such that the drift term \(\mathbf{D}f\) and the collision term \(\mathbf{C}f\) can be treated separately, or subsequently. Hereto, one introduces a discrete time step \(\Delta t\) and discretises the abstract (position) space into cells using enough resolution, such that one can assume spatial homogeneity within each cell \(i\). The collision term is solved for by using a recipe for the particle-particle or particle-medium interactions within each cell, and this term can be used to locally update \(f_t(x_i)\). That new local version of \(f_{t+\Delta t}(x_i)\) is then advected to the next cell via some recipe depending on the local velocity field and a possible external field.

The stochastic, or Monte Carlo, character is introduced by defining the way the collisions or interactions are solved for in each cell, at each subsequent time step. Hereto, one defines a stochastic game transforming the local state vector  in each cell to a newly updated one, the statistical parameters of which can be described by the local cross section coefficients. How this is incorporated depends on the type of transport process. In rarefied gas dynamics, for example, DSMC methods solve for local collisions by defining cross sections based on the number and type of particles present in each cell. As such, the local interaction coefficients are determined by the transported particles themselves. As said, we will concentrate in this paper on linear transport problems, in which the transported particles only interact with a medium, by which the interaction coefficients are determined by the background medium properties only. Note that of course the medium itself might dynamically evolve, by which the cross section coefficients might locally change with time.

\subsection{Particle-medium interactions}

Assuming for now that the medium through which the particles are transported is static, we can elaborate on the general set up as described in the previous subsection. The standard approach to transporting neutrons, electron and photons through such a (possibly active) medium can be described as follows. First, one samples the whole domain onto a rectangular (possibly curvilinear) grid, with enough resolution to consider the medium properties (e.g. the density \(n_i\)) within each cell as constant. Given one or more sources inside or outside of the domain, one determines via some recipe the number or distribution of particles of a certain type originating from one or more grid cells, moving into certain directions. This determines our initial condition \(f(\vec{x}_i,\vec{n},t=0)\), at each grid cell \(i\).

One decouples the drift from the collision part of the transport equation by advecting the particles, or particle distribution \(f(\vec{x}_i,\vec{n},t=0)\), from one grid cell \(i\) to the next \(i'\) in the direction of the particles velocity \(\vec{n}\) during some time \(\Delta t\), whereafter one determines what happens to the particles depending on the medium properties within the new grid cell \(i'\). The local medium properties are characterized by the local cross sections for the various interactions. These can be quite diverse, ranging from scattering to pure absorption. Each of this set of interactions \(\left\{j\right\}\) has a unique coefficient \(\alpha^j(\vec{x}_i)\), which is assumed to be constant throughout each cell.

More specifically, given a total interaction coefficient \(\alpha(\vec{x}_i)=\sum_j \alpha^j(\vec{x}_i)\), one can show that the chance of particles \emph{not} interacting with the medium within that cell is determined by the exponential distribution function

\begin{equation}
  \label{EqChanceSurvival}
  p(s)=\alpha(\vec{x}_i)\mathrm{e}^{-\alpha(\vec{x}_i)s},
\end{equation}
in which one can define \(s\) as the length of the path of the particles from one cell to the next, or the width of one cell. One can randomly sample the probability function Eq.(\ref{EqChanceSurvival}) using a Rejection Method or a Direct Inversion method \cite{NumRecipes}, and determine whether or not particles will interact. What interaction will take place depends on the relative coefficients \(\alpha^j(\vec{x}_i)/\alpha(\vec{x}_i)\). What happens as a result of an interaction depends on the type of interaction. When a particle is absorbed, it is subtracted from the total particle distribution, and when it is scattered, it will be redistributed along a different direction, and advected together with the surviving particles in the next time step. The moments of the distribution function in Eq.(\ref{EqChanceSurvival}) are
\begin{equation}
  \label{EqMFP}
  \left<s^k\right>=\int^\infty_0 s^k p(s)\mathrm{d}s=k!/\alpha^k(\vec{x}_i)=k!\lambda(\vec{x}_i)^k,
\end{equation}
in which \(\lambda(\vec{x}_i)\) is the local mean free path.

Thus, effectively, Monte Carlo methods for the linear transport of particles through a medium, move particles from one interaction to the next, along trajectories which have as an average length the local mean free path, which is determined by the medium properties within each grid cell.

\section{Method description}

In the introduction, we extensively discussed the symmetry-breaking induced by the use of regular lattices, and how this can be resolved by the use of random lattices. In this section, we describe how we combine these random lattices together with the basic ideas of Monte Carlo transport methods to create a new method which solves linear transport equations on an adaptive random lattice.

\subsection{Lattice construction}

We will now proceed with describing how we construct our adaptive, or Lagrangian mesh, along which we will transport the particles. Note that, although we will give examples in 2D, the mesh construction method we will describe in the following is generally applicable in \(d\)-dimensional space for any \(d\ge1\).

\subsubsection{Random lattices}

\begin{figure}
	\includegraphics[width=\columnwidth]{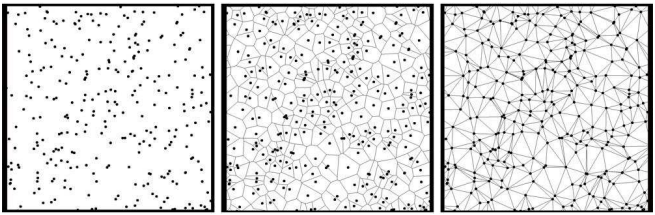} 
	\caption{\label{PoissonFig}\textit{Left}: Result of a Poisson point process. \textit{Middle}:
	The resultant Voronoi diagram. \textit{Right}: The resultant Delaunay lattice.} 
\end{figure}

Standard random lattices are constructed on the basis of Poisson point processes \(\Phi\), which can be defined \cite{Okabe} as the probability 

\begin{equation}
  \label{EqPoisson}
  \Phi=\textrm{Pr}(N(A)=x)=\frac{\left(n_\mathrm{p}\left| A \right|\right) \textrm{e}^{-n_\mathrm{p}\left| A \right|x}}{x!}
\end{equation}
to find \(x=0,1,2,...\) points in any subset \(A\subseteq S=\mathbb{R}^d\). 
The expectation value for Eq.(\ref{EqPoisson}) is \(n_\mathrm{p}\left|A\right|\), in which \(n_\mathrm{p}\)
is the point intensity, which is constant within \(A\). This means that within 
every box of equal volume, the average number of points is equal. An example
of this point process in 2D can be seen in Fig.\ \ref{PoissonFig}, left. The important property
of the Poisson point process is that it can be shown \cite{Okabe} to be
translation and rotation invariant, i.e. homogeneous and isotropic. This makes it an ideal
starting point to construct a lattice which retains these basic properties.

To our knowledge, the least restrictive way to construct a lattice which tessellates
space, and which does not break this rotational symmetry, is the Delaunay triangulation 
\cite{DELAUNAY}. Given a stationary point process \(\Phi\) of nuclei \(\{x_i\}\) in \(\mathbb{R}^d\), 
the Voronoi tessellation \cite{VORONOI} is defined as \(V(\Phi)=\{C_i\}\) in which
\begin{equation}
	C_i=\left\{ y \in \mathbb{R}^d : \left\|x_i-y\right\|\le \left\|x_j-y\right\|
	\forall x_j \ne x_i\right\}.
	\label{Voronoi_Def}
\end{equation}
That is to say, the Voronoi cell \(C_i\) is the set of all points closer to \(x_i\)
than all other points. If two Voronoi cells \(C_i\) and \(C_j\) have a common 
\((d-1)\)-facet (in 2D an edge, in 3D a wall, etc.), they are said to be contiguous
to each other. By joining all the nuclei whose cells are contiguous, we obtain
the Delaunay triangulation, which is a set of simplices and which is, by definition,
dual to the Voronoi tessellation. An example of a Voronoi and Delaunay tessellation 
based on a Poisson point process is depicted in Fig.\ \ref{PoissonFig}.

There are several things to note about this random lattice. First, the
recipe used to construct the Voronoi tessellation, and subsequently the Delaunay
lattice, is solely based on an isotropic distance recipe (cf. Eq. (\ref{Voronoi_Def})). This fact 
ensures the inheritance of rotational symmetry from the point process onto 
the lattice. Second, we are fortunate that specifically for Poisson-Voronoi tessellations there
are several analytically derived properties available \cite{Okabe}. For example, the average
number of edges of a Voronoi cell in 2D, and thus the expected number
of Delaunay lines meeting at one vertex, is equal to \(6\). Thus, on average,
the Poisson-Delaunay lattice is equivalent to the hexagonal lattice used in LB FHP-models \cite{FHP}. In 3D, however, the number of Delaunay lines meeting at one vertex can
be derived to be \((48\pi^2/35)+2\approx 15.54\). The fact that this number is not
an even integer is a direct consequence of the fact that there is no tessellating
Platonic solid in 3D which retains enough rotational symmetry, which as discussed has been known to be a problem in, for example, the LB community. Of course, every individual
Voronoi cell will almost always be asymmetric, but on average it is symmetric.
Thus, we have defined a random lattice which tessellates every dimension \(d\ge 1\), and
is rotationally and translationally symmetric, which makes it ideal for use within those parts of physics where these physical symmetries are absolutely needed. Moreover, spurious invariants associated with the breaking of these symmetries are prevented. As such, these random lattices have been used in the fluid dynamics \cite{Espanol, Flekkoy}, lattice gauge \cite{Christ}, general relativity \cite{Regge} and SUSY \cite{Kaku} communities.

\subsubsection{Lagrangian random lattices}

Although spurious invariant and symmetry breakings associated with rotational and translational invariance are prevented by the use of random lattices, one drawback that still remains is the introduction of an unphysical length scale, determined by the average point to point distance, or the average Delaunay line length \(\left<L\right>\). For Delaunay lattices based on a Poisson point process, the \(k\)-th order expectation value for the line length \(L\) can be derived analytically as \cite{Okabe}

\begin{equation}
  \label{EqExpectedDelaunayLineLengthK}
  \left<L^k \right>=\zeta(k,d)n_\mathrm{p}^{-k/d},
\end{equation}
in which \(\zeta(k,d)\) is some geometrical constant for each pair of the value of \(k\) and the dimension \(d\). Two often used values of this constant are
\begin{eqnarray}
\zeta(1,2) & = & \frac{32}{9\pi} \approx 1.132\label{2DFactor}\\
\zeta(1,3) & = & \frac{1715}{2304}\left( \frac{3}{4\pi} \right)^{1/3} \Gamma\left( \frac{1}{3} \right) \approx 1.237\label{3DFactor}.
\end{eqnarray}

The average Delaunay line length in 3D, for example, is therefore

\begin{equation}
  \label{EqExpectedDelaunayLineLength}
  \left<L\right>=\zeta(1,3)n_\mathrm{p}^{-1/3}\approx 1.237\ n_\mathrm{p}^{-1/3}.
\end{equation}
Although this line length does not have a delta function as a probability function, as in the case of a regular mesh in which the delta function peaks at a length of one cell width, but a certain spread \(\sigma^2=\left<L^2\right>-\left<L\right>^2\propto n_\mathrm{p}^{-2/3}\), it does still have a global first order expectation value that scales with \(n_\mathrm{p}^{-1/d}\), which is a constant once the Poisson point density \(n_\mathrm{p}\) has been chosen. 

As already discussed, these fixed values cause problems when the medium distribution itself is not homogeneous, as they might underresolve high density regions, and impose constraints on parameters like the Knudsen and Reynolds numbers. More generally speaking, it causes problems in regions where the mean free path is shorter than this expectation value.

This problem has been resolved somewhat for regular meshes by introducing AMR, in which the mesh refines itself according to some predefined criterion, often based on the gradient of the pressure or density. Because here we are considering a statistical method, it is justifiably better to choose as an adaption parameter (a function of) the density of the medium. Similar efforts have been made with respect to structured grids, which have one basic congruent cell shape as a basis, in the DSMC community \cite{Garcia, Wu}, in which one wanted to make sure that the local cell sizes at least resolve the local mean free path. In this case, we do not have a regular mesh as a basis and proceed by trying to define a point process which does refine based on the local medium density.

To accomplish this, we discard the Poisson point process, which is the usual basis for random lattices. Instead, we define a point distribution, which is a convolution of a homogeneous Poisson point process \(\Phi\) and a function of the possibly inhomogeneous medium density distribution \(n(\vec{x})\), symbolically written as:
\begin{equation}
  \label{EqPointCorr}
  n_\mathrm{p}(\vec{x})=\Phi * f\left(n(\vec{x})\right).
\end{equation}
The only constraint is the maximum number of points, or resolution, \(N\) available for the simulation. Eq.(\ref{EqPointCorr}) amounts to nothing more than randomly sampling the function \(f\left(n(\vec{x})\right)\) using a Direct Inversion of Rejection method. If the medium distribution \(n(\vec{x})\) is inhomogeneous, one expects the points distribution \(n_\mathrm{p}(\vec{x})\) to be inhomogeneous too, mimicking the medium. But, as long as our number of points \(N\) is high enough, we can always zoom in far enough that locally the medium distribution is homogeneous, and the point distribution Poissonian. Thus, locally, the point distribution defined by Eq.(\ref{EqPointCorr}) retains the rotational and translational symmetries associated with Poisson point processes.

\begin{figure}
\begin{center}
  {\includegraphics[width=4.1cm,clip=]{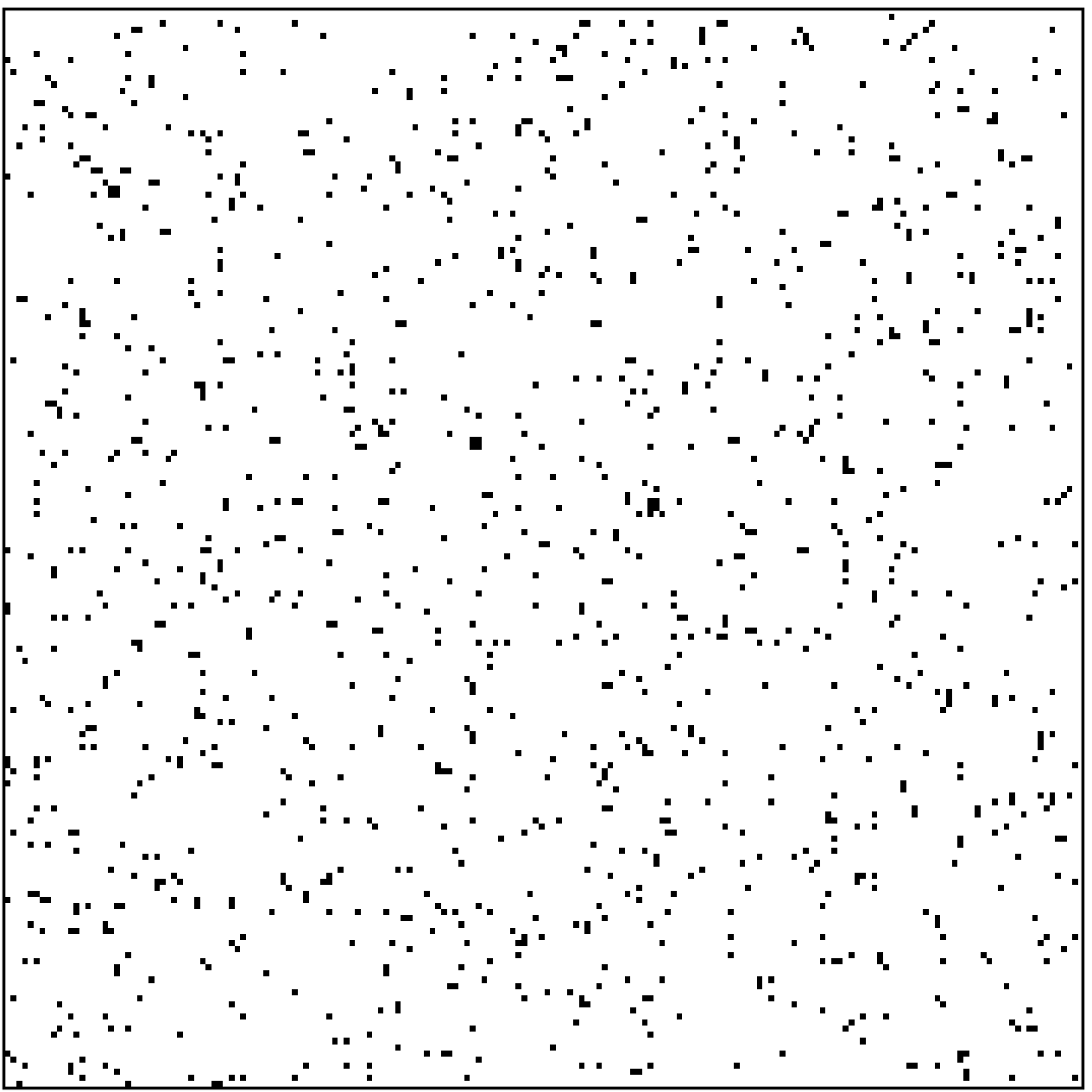}
  \includegraphics[width=4.1cm,clip=]{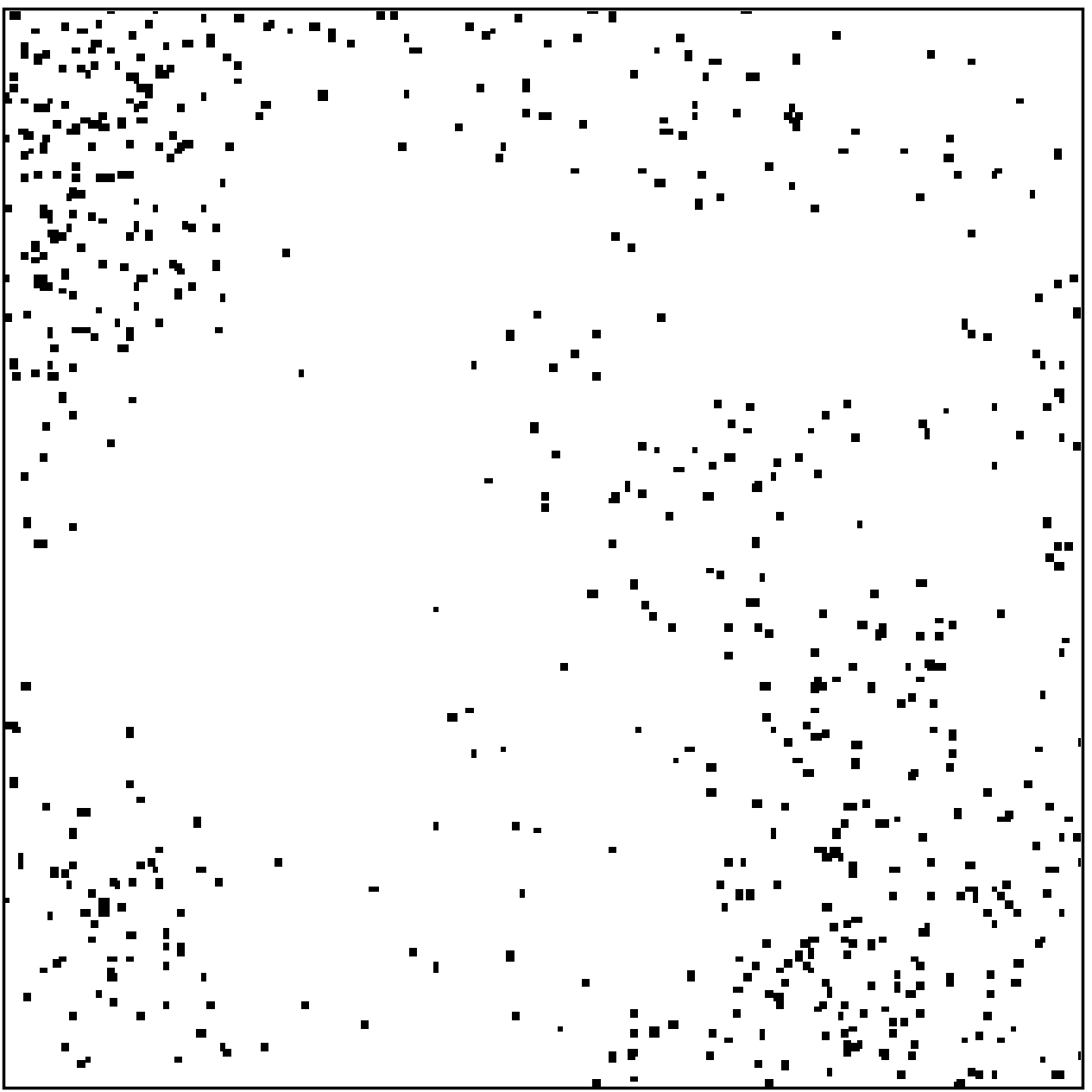}}

  \caption{\label{FigPointDistr}\emph{Left}: Point distribution representing a
  homogeneous medium; \emph{Right}: Point distribution representing a
  clumpy medium.}
\end{center}
\end{figure}

Up until now, we have not specified the exact form of the correlation function \(f(x)\) in Eq.(\ref{EqPointCorr}). We will discuss the details hereof in the next subsection. For now, we will give an example, by choosing \(f(x)=x\), and plotting the resultant point distributions, for two different medium distributions. We refer to Fig.\ \ref{FigPointDistr}, in which is plotted the point distributions for a homogeneous (\emph{left}) and an inhomogeneous (\emph{right}) medium.

\subsubsection{The correlation function}
\label{subsec:CorrelationFunction}

What do we choose for the correlation function \(f(x)\)? Because we introduced this function to ensure adaptation of the point distribution to the medium distribution, the obvious conclusion is that we need \(f(x)\) to be a monotonically increasing function in \(x\), by which the average point-to-point distances will actually be shorter in denser regions.

In light of what we discussed about Monte Carlo methods transporting particles along trajectories which have as an average length one mean free path, we can choose one particular form of \(f(x)\) which  makes the resultant Delaunay line lengths have a very physical meaning.

From basic transfer theory, we know that the local mean free path relates to local medium density in the following way (valid for every dimension \(d\ge1\)):
\begin{equation}
  \label{EqLambdaToDens}
  \lambda(\vec{x})=\frac{1}{n(\vec{x})\sigma},
\end{equation}
in which \(\sigma=\sum_j \sigma_j\) is the total cross section, possibly consisting of many different interaction cross sections \(\sigma_j\), each having its own mean free path \(\lambda_j=1/n(\vec{x})\sigma_j\). Because the mean free path is a statistical length, it scales in a different way with the medium density than the average Delaunay line length, which has an extra dimension dependence [cf. Eq.(\ref{EqExpectedDelaunayLineLengthK})]. From that, we can easily conclude that, if we choose our point distribution to sample the \(d\)-th power of the density, i.e. \(f(x) \propto x^d\), or, more specifically,
\begin{equation}
  \label{EqPointDensToDens}
  n_{\mathrm{p}}(\vec{x})=\Phi * n^d(\vec{x}), 
\end{equation}
the length of a Delaunay
line \( \left<L\right>(\vec{x}) \) between two points, will scale linearly with the \emph{local} mean
free path of the medium \( \lambda(\vec{x}) \) via a constant \(c\). That is
\begin{equation}
  \label{EqLengthToMFP}
  \left<L\right>(\vec{x}) = c\lambda(\vec{x}).
\end{equation}
Thus, because we choose the point distribution to conform to the density profile of the medium
according to Eq.(\ref{EqPointDensToDens}), the average Delaunay line length and the mean free path
have the same \(n^{-1}\) dependence, by which Eq.(\ref{EqLengthToMFP}) is a \emph{global} relation with a \emph{global} constant \(c\).

There are two things to note. First, the medium, and thus the medium density distribution, might evolve. In that case the lattice, which is Lagrangian by definition of Eq.(\ref{EqPointDensToDens}), will evolve with it. In most relevant cases, the transport of particles through a medium is studied with respect to static media, but for several cases, such as radiation hydrodynamics, it is worthwhile to keep in mind that the medium density \(n(\vec{x})\), and thus the point density \(n_\mathrm{p}(\vec{x})\), can change with time, according to what a separate hydro solver provides us with. This is also how we see a possible extension to the nonlinear transport regime of, for example, hydrodynamics. As in the DSMC method, we could define a linear Monte Carlo-like transport process through a background medium, now consisting of the particles themselves; in this case, however, the background medium, i.e.\ the particles themselves, can not be considered static. As such, the random lattice needs to be updated very frequently, which is a costly, and intricate task.

Second, one can choose the correlation function to be any monotonically increasing function different from the one in Eq.(\ref{EqPointDensToDens}), but in that case the global constant \(c\) would change into a locally varying function \(c(\vec{x})\). For example, if we choose \(f(x)\propto x^e\), where \(e\ge0\), that varying function would be \(c(\vec{x})\propto n^{(d-e)/d}\).

\subsection{Lattice properties}

In the previous subsection, we described how we construct the adaptive random lattices based on the medium density distribution. Before we set out to define the way one can transport particles on this lattice, we need to discuss the exact statistical properties of the lattice, and the errors associated with its stochastic nature.

\subsubsection{Distributional equivalence}

The linear correlation of Eq.(\ref{EqLengthToMFP}) can be taken one step further, by relating Eq.(\ref{EqMFP}) to Eq.(\ref{EqExpectedDelaunayLineLengthK}), and recognizing that, by choosing a correlation function \(f(x)\propto x^d\),  not only the first order moment, but \emph{all} \(k\)-th moments of the exponential distribution in Eq.(\ref{EqChanceSurvival}) will scale linearly with the \(k\)-th order expectation of the Delaunay line length, i.e.

\begin{equation}
  \label{EqScaleAllOrders}
  \left<L^k\right>(\vec{x})=c(k)\lambda^k (\vec{x}),
\end{equation}
in which \(c(k)\) is still a global constant, but one that now depends on the order \(k\). This similarity in statistical properties is, of course, not very surprising, because the interval between two events, or the distance between two points, in the Poisson point process in Eq.(\ref{EqPoisson}) has an exponential distribution \(p(x)=n_\mathrm{p}\left|A\right|\mathrm{e}^{n_\mathrm{p}\left|A\right|x}\) equivalent to that of the path length between two events in Eq.(\ref{EqChanceSurvival}). It is to be expected that the distribution function for the average distance, or Delaunay line lengths between two of those points follows a similar distribution, with some modifications because of the dimension of space. The mean free path statistical length is one-dimensional always, not depending on dimension, so by choosing the correlation Eq.(\ref{EqPointDensToDens}) we remove the dimensional dependence of the expectation values Eq.(\ref{EqExpectedDelaunayLineLengthK}), by which the path lengths of particles \emph{and} the average lattice line lengths are distributed similarly.

\subsubsection{Length sampling}

We described how to construct a lattice, which is homogeneous and isotropic locally, and which adapts to the medium properties. Moreover, we showed that, by choosing a smart correlation function, the lattice becomes a direct representation of the `free path space' of the particles. That is, locally, the Delaunay line lengths have the same distributional properties, i.e. the same \(k\)-th order moments, as the path lengths until interaction of the particles.

Thus, the Delaunay line originating from one point in the medium has distributional properties  (\(k\)-th order moments) which all scale linearly with the distributional properties of the path lengths originating from that point. Stated differently, the variance of the Poisson process is now not associated with noise, but is exactly the variance of the free path of the particle.

Given a number of points, or resolution, \(N\), we construct one instance of the ensemble of the point distribution, and the free paths are accurately sampled at each one of those points. This could cause inaccuracies, once the medium between two points in not locally homogeneous, by which it does not have enough sampling points. Thus, we need to impose as a sampling condition, that in the vicinity of each point (or, more accurately, within each Voronoi cell), the medium can be considered homogeneous, a condition needed in almost every numerical method. Symbolically, we need \(\frac{1}{n(\vec{x})}\frac{\partial n(\vec{x})}{\partial x}<\frac{1}{\left<L\right>}\).

This can be dealt with in two ways. First, one can increase the number of points \(N\), by which the global parameters \(c(k)\) decrease correspondingly, until the condition is satisfied. It is obvious that, when \(N\rightarrow \infty\), the density field is sampled continuously, and the result is exact. Second, when the available resolution is less than needed, we can construct several instances of the same point distribution, and overlay them afterwards. This is allowed, because the expontential distribution, on which the Poisson point process is based, is memoryless, by which the individual instances are independent. Note that the sampling condition is reached faster by choosing a correlation function as in Eq.(\ref{EqPointDensToDens}). 

\subsubsection{Angular sampling}
\label{subsec:AngularResolution}

Another variable which has to be sampled accurately, is the number of directions a particle can propagate into at each point. As said, the average number of directions at each point is fixed (\(6\) and \(15.54\), in 2D and 3D, respectively), and will not increase, when the resolution or the number of instances increases. We have to take this into account, when we design a transport algorithm, in which we want to conserve momentum at each point. Moment conservation can be imposed, as we will show in the next subsection.

If we construct many instances of the ensemble of point distributions, the directions of the lines will differ for each instance, and, eventually, the continuous rotation group \(SO(d)\) will be sampled continuously. We can, however, apply the ergodic principle, and state that it is equivalent to replace the ensemble average, by a volume average. The number of directions within a volume, containing \(N\) points, scales as \(O(N)\). Thus, within a certain locally homogeneous medium, the number of points, and henceforth the number of lines has to be large enough to accurately sample the unit sphere. This amount to nothing different to stating that a random lattice is isotropic.

Thus, we can conclude that, when the number of points within a certain small region will increase towards infinity, the number of directions will follow that trend, and the angular sampling will become infinitely precise.

\subsection{Transport}

The Lagrangian random lattice, described in the previous subsection, is a direct representation of the `free path space' of the medium. As such, the stochastic element of the linear transport of particles on a fixed, deterministic grid in regular Monte Carlo methods, is moved to the simple deterministic transport of particles on a lattice, which now has the stochastic properties. Moreover, we can dispense with the regular grid or the underlying medium altogether, because all the information needed for the transport of particles has been translated to the lattice, by which the transport of particles through a medium has been translated to the transport, or percolation, of particles on the graph consisting of the lattice lines.

In the following, we will describe how we can use this lattice as a basis for transporting particles, depending on what kind of process we would like to model. As an example, we will use the two elementary linear transport processes, ballistic transport (possibly with an additional absorption term) and transport through a scattering medium, and combinations hereof.

\subsubsection{Ballistic transport}
\label{subsec:BallisticTransport}

We will commence with describing how we can transport particles along the lattice for a medium in which the scattering cross section is negligible. Because, in this case, we transport particles which have a predetermined momentum, it is of utmost importance to define what we do at each grid point to ensure momentum conservation, and prevent numerical diffusion.

Because our method works in such a way that the line lengths correlate linearly with the mean free paths, we assume that the homogeneously distributed medium is absorptive, and that the associated cross section \(\sigma_\mathrm{abs}\) is the only contribution to the total cross section. Henceforth, we obtain a lattice which is similar to the Delaunay graph in Fig.\ \ref{FigPointDistr}, right.

We define one point as a source of particles, sending them out along one of the lines. The particles move along the line, until they come upon the next point. They have moved along a line, which correlates linearly with the mean free path via a constant factor \(c_\mathrm{abs}\). We can exactly evaluate the value of this factor, given Eq.(\ref{EqExpectedDelaunayLineLengthK}) and Eq.(\ref{EqPointDensToDens}), as

\begin{equation}
  \label{EqCoeff}
  c_\mathrm{abs}=\xi(d,N,D)\sigma_\mathrm{abs},
\end{equation}
in which \(\xi(d,N,D)\) is some constant depending on the choice of dimension \(d\), the number of points \(N\), and of the size of the domain \(D\). Note that each other type of particle-medium interaction \(j\) would have a coefficient \(c_j\) determined via a relation similar to that in Eq.(\ref{EqCoeff}).

Ideally, one would want all \(c_j=1\), but, almost always, the resolution \(N\) will result in a coefficient larger than unity, in which case the line length is larger than the local mean free path, or smaller than unity, in which case it is the other way around. This can be taken care of be defining how the interaction, in this case absorption, is accounted for at each point.

Hereto, we define the incoming number of particles, or intensity, as \(I_\mathrm{in}\), and determine the outgoing intensity as

\begin{equation}
  \label{EqAbs}
  I_\mathrm{out}=I_\mathrm{in}\mathrm{e}^{-c_j},
\end{equation}
which is equivalent to the familiar \(I'(x)=I_0\mathrm{e}^{-x/\lambda}\), which can be derived from Eq.(\ref{EqChanceSurvival}). Note that, with \(c_j\) being a global constant, \(\mathrm{e}^{-c_j}\) is a global constant too, which can be determined a priori, via Eq.(\ref{EqCoeff}). In some cases, when the resolution \(N\) is high, and thus the coefficients \(c_j\) small, it might be useful to approximate \(\mathrm{e}^{-c_j}\approx (1-c_j)\).

Locally the number of particles absorbed can be exactly evaluated as

\begin{equation}
  \label{EqRetained}
  I_\mathrm{abs}=I_\mathrm{in}\left(1-\mathrm{e}^{-c_j}\right),
\end{equation}
which ensures that this method conserves particles exactly.

The remaining particles \(I_\mathrm{out}\) have to be sent out along one of the lines emanating from this point. The directions we choose depends on whether we want to conserve momentum, or if the interaction was with a scattering medium, in which case we want to distribute the particles isotropically. In the present case, we need to ensure momentum conservation, by which the we have to choose as an outgoing line one which is in the same direction as the incoming one.

As we already discussed in the previous subsection, the lines only sample the unit sphere on average, and although the Voronoi cells are cylindrically symmetric, on average, with respect to every incoming Delaunay line, every one particular cell will deviate from that. This has as a result, that almost always, there is no outgoing line in the same direction as the incoming one. Thus, particles are deflected from their original direction by the irregularity of the grid. This can be viewed as the introduction of space-dependent inertia forces, and it is very important to keep track of these, especially in the nonlinear regime \cite{Benzi, Karlin}, which we do not consider in this paper.

\begin{figure}
	\includegraphics[width=0.7\columnwidth]{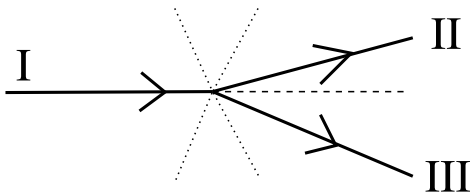}
	\caption{\label{FHP_fig}Example of a number of Delaunay lines meeting at
	a node. Incoming particles (along line I) interact with the medium at the node,
	and the remaining particles should continue along the dashed line to conserve momentum. Choosing 	lines
	II and III as outgoing ensures conservation of momentum, on average.}
\end{figure}
We can resolve this by doing the following. We refer to Fig.\ \ref{FHP_fig}, which is an example in 2D space. In this Figure, the dashed line is the line along which momentum would be conserved.
Instead, if we now choose the `nearest' line II as the outgoing line, in which nearest can be defined
in several ways, the most intuitive being that line for which the inner product with
the dashed line is largest, it can be shown that momentum is conserved on average,
which immediately follows from the fact that the Voronoi cell is 
axisymmetric with respect to every incoming Delaunay line.
However, from a numerical implementational point of view, it is more elegant to
split up the two particles and distribute one of them along line II and the other
along line III, being the next to most straightforward path. More generally,
if the transported quantity is a continuous entity, we have found it is most 
efficient to split this quantity into \(d\) equal parts, in which \(d\) is the dimension
of our computational domain, and distribute each of these parts along the \(d\)
most straightforward paths. Because by definition, the Poisson-Voronoi cells
are axisymmetric with respect to every associated Delaunay line in every dimension
\(d\ge 2\), we know that a similarly modified set of rules will ensure
conservation of momentum.

\begin{figure}
	\includegraphics[width=\columnwidth]{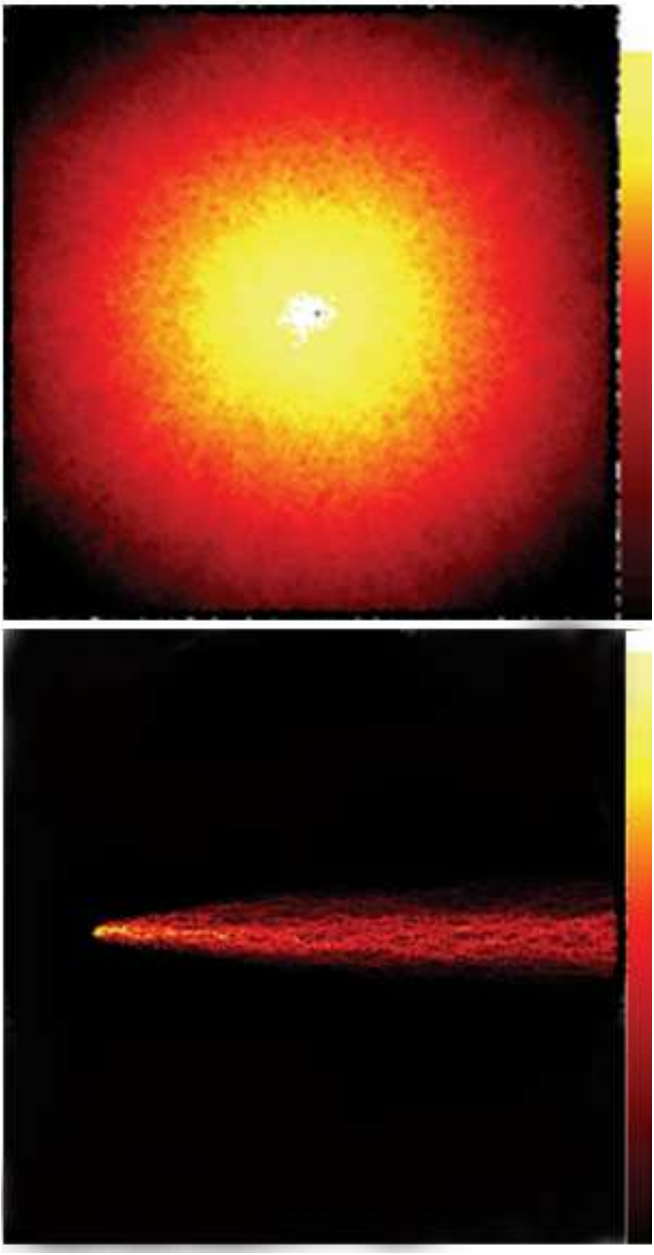}
	\caption{\label{FigExamples}(Color online) The result of two simple test on a 2D Poisson-Delaunay
	random lattice with \(N=5\cdot 10^4\) points. Both are 
	logarithmic plots of the number of particles at each site. \textit{Top}: illustration
	of a scattering transport process; 
	\textit{Bottom}: illustration of the conservation of momentum by means of the
	transport of particles with constant momentum vectors.}
\end{figure}

To demonstrate conservation of momentum, we constructed a 2D random lattice of a Poisson point process with a relatively resolution of \(N=5\cdot 10^4\) points, which is sufficient to show the trends and is still coarse enough to clearly show noise.
Next, we define one point as source of particles, each
emitted with the same momentum vector. If the particles were defined to be photons,
this source could for example be called a laser beam. The rules at each
site are chosen such that the two most straightforward paths with respect to this
momentum vector are chosen, and that the incident package of particles are split
in two and continue along these two lines. We follow the particles until they hit
the absorbing boundary. The result is plotten in the bottom half of Fig.\ \ref{FigExamples},
in which we plotted the logarithm of the number of particles at each vertex, given
that the source emits a high number of particles. 

One immediately sees that, on average,
the resultant momentum vector is in the original direction (by the symmetry properties of the Delaunay lattice \cite{Okabe}), but that there is some
inherent noise associated with the use of these random lattices. Fortunately, one can prove that this noise vanishes in the limit of \(N\rightarrow\infty\), by
recognizing that in \(d\)-dimensional space the propagation of a particle on a lattice with this set of rules is equivalent to the process of an anisotropic random walk on a graph. An exact mathematical derivation is given in Appendix \ref{app:ConservationOfMomentum}, and we give a simplified version here.

Defining a signal \(\mathcal{S}\propto \left<L\right>n\) as the (average) distance travelled after \(n\) steps,
and realizing that for an anisotropic random walk we have a noise
\(\mathcal{N}\propto\left<L\right>\sqrt{n}\), we obtain a signal to noise ratio \(\mathcal{S}/\mathcal{N}\propto
n^{1/2}\) (which is of course similar to the famous inverse square root law of Monte Carlo methods). This is why the beam in Fig.\ \ref{FigExamples} does not diverge. Given
a particle location \(x\) along the momentum vector at a distance \(s\) from the source, 
we know that the average number of steps \(n\) for a particle to reach \(s\) scales as
\(n\propto s/\left<L\right>\) in which \(\left<L\right>\propto N^{-1/d}\). From this, we can conclude that
the signal to noise ratio at a distance \(s\) from the source scales as
\begin{equation}
	\frac{\mathcal{S}}{\mathcal{N}}\propto N^{1/2d}.
	\label{SignalNoise}
\end{equation}
Thus, in the limit of \(N\rightarrow\infty\), momentum is conserved exactly, not
just on average. In other words, the width of the beam will shrink to zero. Note that, when \(N\) increases, the interaction coefficients \(\{c_j\}\) will decrease. 

Thus, we have shown that choosing this transport recipe for ballistic transport will not only conserve momentum on average, but it does also deal with the coarse angular sampling at each point of the lattice, ensuring that any numerical diffusion will be minimal. Moreover, we have shown that this artificial widening of the beam will go to zero for infinite resolution.

\subsubsection{Transport through scattering media}
\label{subsec:Scattering}

In the previous subsection we discussed, how to do transport of particles through a medium for which the interaction conserves momentum, and does not change the original momentum vectors of the particles. It is also possible that one of the interactions \(\{c_j\}\) can also influence the vectorial properties of the particles, for example when the transport is through a scattering medium.

Without loss of generality, we assume a similar homogeneous medium as before, but now, without absorptive properties, but \emph{with} an extra interaction coefficient, \(\sigma_\mathrm{scat}\), which accounts for the scattering properties of the medium. In this case, the method for transportation is rather similar as in the previous subsection, in that we propagate the scattered particles \(I_\mathrm{out}=I_\mathrm{in}\mathrm{e}^{-c_\mathrm{scat}}\) along the \(d\) most straightforward paths. What differs, is that the retained particles \(I_\mathrm{ret}=I_\mathrm{in}\left(1-\mathrm{e}^{-c_\mathrm{scat}}\right)\)
will now be redistributed isotropically. Hereto, one can choose to propagate an equal section along every one Delaunay line. Of course, the angular sampling is not perfect, but similar considerations as made in Sect.\ \ref{subsec:AngularResolution} will ensure angular resolution.

An example of one such experiment is depicted in Fig.\ \ref{FigExamples}, top, in which we define several point is a small central region as sources, and for which we let the particles propagate along the lattice according to the rules described above. We take several points as a source in this case, because this will ensure that we have many different original directions. Note that the result is very isotropic, as is to be expected of this random lattice. Moreover, the result exactly depicts the noise associated with the finite angular sampling, and thus the finite amount of points.

\subsubsection{General interactions}

The particle-medium interaction can consist of many different types, each one having its own contribution to the right side of Eq.(\ref{EqAbstractBoltzmann}). Each one of these interactions has its own cross section \(\sigma_j\), and its own corresponding linear correlation coefficient \(c_j\). Note, however, that the conversion factor \(\xi(d,N,D)\) in Eq.(\ref{EqCoeff}) is the same for all interactions, given one choice of \(d\), \(N\) and \(D\).

The total interaction factor \(C=\sum_j c_j\) determines what factor of the incoming particles interacts, in one way or the other, \(I_\mathrm{out}=I_\mathrm{in}\mathrm{e}^{-C}\). Every single interaction in the set \(\{c_j\}\) has its own contribution \(\frac{c_j}{C}\mathrm{e}^{-C}\). When our resolution is conveniently high enough, and thus our factors \(\{c_j\}\) correspondingly small, we can approximate the total factor of the particles that interact, as \(\left(1-\mathrm{e}^{-C}\right)\approx C\), by which one can conclude that every interaction retains a factor \(c_j\) of the incoming particles. This last approximation can be implemented more efficiently code-wise.

What happens to particles, when they interact, of course depends on the type of interaction, and the type of transport process in general. Particles can be redistributed isotropically, and continue on through the medium, or they can influence the medium, by heating, ionization, or something similar. They might even be re-emitted as different particles. All these type of interactions can be implemented very easily, and the feedback of the particles on the medium is defined straightforwardly. One might even have multiple species of particles, in which case there would be a cross section \(\sigma_j\) for each type of particle, and the set of factors \({c_j}\) would change into a matrix, which could include the transformation of one particle into the other.

We conclude, by noting that one needs to aim at having enough resolution to ensure \(c_j\le1\). In cases, where scattering can be neglected, the solution based on Eq.(\ref{EqAbs}) would still be exact, as we could not resolve the space between the points anyway, but when wanting to incorporate scattering, one might not accurately resolve the diffusion coefficient, when \(c_\mathrm{scat}\gg1\).

\subsection{Time stepping}

Up until now, we have not elaborated on what we define as a time step. As in linear Monte Carlo methods, how to define the time step depends largely on the transport problem at hand. We can mostly define two distinct cases: A) the interaction dominated limit (\(\lambda/L\ll 1\)), and B) the free streaming limit (\(\lambda/L\ge 1\)). 

The first case is what is mostly encountered in linear transport problems. Here, the mean free paths for the particles are so small compared to the size of the domain, that we know the particle will be absorbed somewhere within the domain. Given that the speed of the particles is very high, one can, for each time step, let the particles that are emitted in that interval move along the lattice, moving from one interaction point to the next, until they are annihilated. The time steps do have to be chosen such that one can accurately sample time dependent source functions, or such that, when absorption is followed by ionization (cf.\ Sect.\ \ref{subsec:Testcase}), one can accurately follow the ionization front. Equilibrium solutions can be found, if they exist, by having a constant number of particles being emitted at each time step, that will exactly compensate for the number of particles that are absorbed, or leave the computational domain.

The second case is not very interesting from an particle-medium interaction point of view, because the densities and cross sections are of such form that no interactions are expected to occur. Thus, the particles can stream freely through quasi-vacuous space. In this case, one may be interested in following the particles' trajectories, and the time steps are then dictated by the size of the domain divided by the speed of the particles.

An interesting, albeit a bit artificial, intermediate case is the one described in Sect.\ \ref{subsec:Scattering}. In this case of pure scattering, the particles are not expected to be annihilated. If we would not stop the particles at one point, they would eventually leave the domain. Thus, if we want to, for example, accurately follow the spherical wavefront around a point source in a homogeneous, scattering medium, as it expands with time according to the familiar Gaussian diffusion profiles, we need to keep a clock for each particle, such that it does not take more mean free path steps as is allowed by, for example, its maximum speed.

It is clear that the relevant time step criteria depend greatly on what it is one is trying to solve for. In each case, however, one can define the relevant time stepping unambiguously. Most linear transport problems we are interested in, including the one discussed in Sect.\ \ref{subsec:Testcase}, will be of type A.

\subsection{3D and beyond}

Because it may not seem obvious how our method is trivially extendible to \(\mathbb{R}^d\), given all the 2D examples we gave, we will explicitly state how this extension is automatically achieved by the lattice construction procedures we gave.

First, the recipe for constructing our adaptive point process using Eq.(\ref{EqPointDensToDens}) is applicable and valid in general dimensional space. Second, the procedure for constructing the Delaunay triangulation, and the corresponding recipe for the Voronoi tessellation Eq.(\ref{Voronoi_Def}), can be used and implemented efficiently in every \(\mathbb{R}^d\) \cite{Okabe}. Because we chose a correlation function that includes the dimension of space \(n_\mathrm{p}\propto n^d\), we made sure that the linear relation between all line lengths of the resultant lattice and the local mean free paths is valid in every dimension, cf.\ Eq.(\ref{EqScaleAllOrders}). Thus, we have obtained an adaptive random lattice that adapts to the medium in exactly the same for for every \(\mathbb{R}^d\).

As we have already pointed out, the resultant transport process is nothing more than a walk from on interaction event to the next (as it is in Monte Carlo methods), or, alternatively, a walk from one vertex to the next along a \(d\)-dimensional Delaunay graph that has mean free path-like line lengths. The recipes for the different types of processes (ballistic, scattering, and combinations) can be extended trivially to \(\mathbb{R}^d\). For example, for ballistic transport, we do not choose the two most straightforward paths, but the \(d\) most straightforward. The analysis concerning the conservation of momentum in the appendix is valid for every dimension, because it only assumes that the typical Voronoi cell is cylindrically symmetric around every associated Delaunay edge. This symmetry is a natural consequence of the motion invariant property of the underlying Poisson point process (see \cite{Okabe}, for more details).

The only things that change from dimension to dimension are the average number of lines emanating from a typical point, or the number of walls of a typical Voronoi cell (\(6\) in \(\mathbb{R}^2\), \(15.54\) in \(\mathbb{R}^3\), etc.), and the geometrical constant \(\zeta(k,d)\) in the expressions for the \(k\)-th order moments of the Delaunay line lengths Eq.(\ref{EqExpectedDelaunayLineLengthK}). These are mere constants that one has to incorporate when implementing the numerical method.

When using a regular grid, and the often associated finite-differencing, or some nontrivial form of interpolation, it is often very nontrivial to extend these operations to higher dimensional space. Because we chose to use an adaptive random lattice on which a (random) walk is performed, these difficulties are resolved, because both the lattice construction techniques and the (random) walks are trivially defined in every \(\mathbb{R}^d\).

\subsection{Efficiency}

It is straightforward to implement our method using one or the other programming language. We have already done this, using \texttt{C++}, and we will describe an application of this so-called \texttt{SimpleX} package in the next section.

For now, we will describe the basic steps of the algorithm. First of all, there have to be several pre-processing steps: 1) create a point process matching Eq.(\ref{EqPointDensToDens}), in which the medium density function can be an analytic function or some data array from some other simulation; 2) construct the Delaunay triangulation; 3) determine the properties of the resultant lattice, namely the global interaction coefficients \(\{c_j\}\), and the \(d\) most straightforward paths with respect to the other lines. All these values are fixed during the rest of the simulation. As an illustration, given an efficient tessellation code, these preprocessing steps can be completed within one minute wall-clock time on a simple desktop computer, for a resolution of \(N=10^6\) points.

Once the lattice and all its properties are known, the transport can commence. If we define one iteration of the algorithm as advancing particles from one point to the next, and subsequently performing the interactions at each point, and that for each point, we can make an estimate of the operation count of the algorithm. 

Each point can be dealt with independently of all the others (cf. the Markov property of Monte Carlo methods). Given that the particles can be redistributed along other angular directions at each point, each line has to treated separately, but this is just a geometrical constant, given the dimension \(d\), which does not scale with any resolution. Linear multiplications as in Eq.(\ref{EqAbs}) have to be performed for each interaction \(j\), and for each different particle species. The redistribution along the lines is just a pointer operation. Thus, the total operation count is of the form
\begin{equation}
  N_\mathrm{ops}=N_\mathrm{int}N_\mathrm{spec}N_\mathrm{p},
\end{equation}
in which \(N_\mathrm{int}\), is the number of different interactions or non-negligible cross sections \(\sigma_j\), \(N_\mathrm{spec}\) is the number of species, and \(N_\mathrm{p}\) is just the resolution, or number of points. Thus, just focusing on the resolution, the operation count of this method scales as \(O\left(N\right)\). 

It should be noted that this is \emph{in}dependent of the number of sources. As it turns out, most other transport methods scale with the number of sources, by which it is extremely time-consuming to do realistic calculations for a large number of inhomogeneously distributed sources. With this method, this is now feasible, even on a simple desktop computer, as we will point out in the next section.

We end this section by pointing out that our method bears much resemblance to cellular automata methods, in the sense that we have a global set of interaction coefficients, and a global set of rules, applied locally. Each point is influenced only by its neighbors (via the Markov criterion). Thus, like most cellular automata methods, our method can be parallelized trivially. Of course, this would be much less trivial, when nonlinear terms would be included.

\subsection{Concluding}

In this section, we have described how our new method works. We have shown how to construct a Lagrangian random lattice, which mimics the medium properties in such a way that locally all line lengths have the same distributional properties as the particles' path lengths. As such, the method can handle any geometry of the medium. The grid retains the translational and rotational symmetries, inherent to most physical problems, whilst at the same time adapting to the medium properties, by which small mean free paths and rapidly fluctuating regions of the medium will not be undersampled. The resultant lattice is a direct representation of the free path distribution space of the medium, by which the stochastic character of the the particles' trajectories as in regular Monte Carlo has been lifted to that of the grid itself, which can be shown to sample space, and all angular directions exactly when the resolution goes to infinity. Particles can be easily transported along the lattice lines, with the interactions taking place at each grid point, via an interaction coefficient \(c_j\), which is directly proportional to the interaction cross section. We have discussed how time-stepping is introduced, depending on the linear transport problem we try to solve for, and we have demonstrated how the often not very trivial extension from \(\mathbb{R}^2\) to \(\mathbb{R}^d\) \emph{is} trivial for our method. Moreover, we have shown that the operation count of an implementation of the method is \(O\left(N\right)\), which makes the method fast, even when increasing resolution, or performing 3D calculations.

\section{Radiation Transport}

As already briefly touched upon in the previous section, we have implemented the method described in this paper in a \texttt{C++} package, we called \texttt{SimpleX}, named after the elementary constituent of the Delaunay lattice. For the grid construction phase, we make use of the open source package \texttt{QHull} \cite{QHULL}, which has been proven to be as fast as the mathematically
established limit \(O(N\log N)\) for \(d\le3\) and \(O\left(N^{\lfloor d/2 \rfloor}\right)\)
for \(d\ge4\).

The \texttt{SimpleX} package was specifically designed to solve the equations of radiation transport. This radiative transfer comes in a wide variety of forms, depending on the (astro-)physical applications one is looking at. In the following, we will concentrate on one particular example \texttt{SimpleX} has been used for, and that is the propagation of ionizing photon in the Early Universe.

\subsection{Reionization}

When the Universe had an age of about \(400,000\) years, its expansion had caused it to cool down such that the hydrogen recombination rate was higher than the ionization rate, by which almost all protons and electrons recombined into neutral hydrogen, making the Universe opaque to ionizing radiation. It took quite some time, before the initial density perturbations gave rise to the first generation of stars (dubbed \emph{population III}) and quasars, which could end these Dark Ages by producing the first new supply of ionizing photons capable of reionizing the neutral hydrogen. This period, which starts with the first new photons being produced, and ends when all the hydrogen has been ionized, is believed to have happened in the redshift span \(6<z<20\) (i.e. \(150\) million - one billion years after the Big Bang), and has the telltale name of \emph{Epoch of Reionization} (EOR) \cite{Loeb, Bromm, Ciardi}. This part of cosmology has received considerable attention in recent years, because we believe to be at the verge of observing signatures of the EOR itself, and understanding what physically happens when reionization starts is absolutely mandatory.

The initial density perturbations will enhance and enter a stage of non-linear growth, until, at the dawn of the EOR, dark and gaseous matter is distributed along a very inhomogeneous filamentary structure, in which the density range between high density filaments and low density voids can span many orders of magnitude. The photon consumption is dominated by the small scale structure, which one therefore needs to resolve, but the resultant ionization bubbles can have sizes which span sizes many magnitudes larger. It is therefore of the utmost important for methods trying to model the EOR to have large enough simulation boxes in combination with a resolution range large enough to be able to resolve the small scale structure. This, together with the inhomogeneous distribution of the matter, \emph{and} the sources, makes it an ideal situation where the Lagrangian aspect of our method is aptly suited. Moreover, the large number of sources involved slow other methods down severely, sometimes even beyond the reach of modern supercomputers. The \texttt{SimpleX} method does not have this scaling property, and can therefore be easily used, even in these extreme cases. 

\subsection{Setup}

In this specific case, the medium density \(n(\vec{x})\) is determined by the hydrogen density \(n_\mathrm{H}(\vec{x})\), which is provided by dark matter simulations which let the matter distribution evolve from initial perturbations in the microwave background radiation up until the beginning of the EOR. Our code translates this density distribution into a point distribution via Eq.(\ref{EqPointDensToDens}). The standalone hydro-code also provides us with a catalogue of where the first sources have formed, together with their intensities, conveniently expressed in terms of number of ionizing photons emitted per second.

Each point is designated the same global interaction coefficient \(c_\mathrm{ion}\), associated with the photo-ionization interaction, but we include an extra factor \(\chi \in [0,1]\), which is a parameter that accounts for the feedback to the medium, and keeps track of what factor of the local medium is still neutral. This neutral fraction is updated after each iteration, and, effectively, lowers the interaction coefficient, i.e.
\begin{equation}
  \label{EqEffIon}
  c_\mathrm{eff}=\chi c_\mathrm{ion}.
\end{equation}

An extra time-dependent effect, that is needed to slow down the resultant ionization fronts, is hydrogen recombination, which can be locally incorporated by evaluating at each point, at the end of each iteration,
\begin{equation}
  \label{EqRecomb}
  \dot{n}_\mathrm{rec}(\vec{x})=\alpha_\mathrm{B}(1-\chi)^2 n^2_\mathrm{H} (\vec{x}),
\end{equation}
in which \(\dot{n}_\mathrm{rec}(\vec{x})\) is the local recombination rate and \(\alpha_\mathrm{B}\) is the recombination coefficient. This expression effectively ignores diffuse, or scattered, radiation, which is thought to be unimportant, although recent doubt has been shed on the validity thereof \cite{Ritzerveld}. Note that, if diffuse radiation turns out to actually be a real important factor, the method described in the previous section can trivially include it.

There are three things to note. First, the time-step \(\Delta t\) per iteration is determined by how much photons we emit per iteration, and can be made arbitrarily small, in order to accurately resolve effects like Eq.(\ref{EqRecomb}). Second, we can built in a limiter at each point, to make sure the velocity of the ionization front will not exceed the speed of light. And, third, in this case of radiation transport, it is extremely important that we have defined our method to be particle conserving, because precisely this ensures the ionization fronts to have the correct speed.

\subsection{Testcase}
\label{subsec:Testcase}

There is one classical problem, which serves as an excellent testcase for all cosmological radiation transport code, and that is the one of an ionization front expanding in an initially neutral and uniform medium \cite{Stromgren, Spitzer}. It so happens that the exact analytical time-dependent solutions are known for this problem.

Given that the medium has a homogeneous density \(n_\mathrm{H}\), and a central monochromatic source emits photons with energy \(h\nu=13.6\ \mathrm{eV}\) at a rate \(\dot{N}_\gamma\), the solutions for position \(r_\mathrm{I}(t)\) and velocity \(v_\mathrm{I}(t)\) of the ionization front are
\begin{eqnarray}
  \label{EqIonFrPos}
  r_\mathrm{I}(t) & = & r_\mathrm{S}\left(1-\mathrm{e}^{-t/t_\mathrm{rec}}\right)^{1/3} \\
  \label{EqIonFrVel}v_\mathrm{I}(t) & = & \frac{r_\mathrm{S}}{3t_\mathrm{rec}}\frac{\mathrm{e}^{-t/t_\mathrm{rec}}}{\left(1-\mathrm{e}^{-t/t_\mathrm{rec}}\right)^{2/3}},
\end{eqnarray}
where
\begin{equation}
  \label{EqRecombTime}
  t_\mathrm{rec}=\frac{1}{\alpha_\mathrm{B}n_\mathrm{H}}
\end{equation}
is the local recombination time and
\begin{equation}
  \label{EqStromgren}
  r_\mathrm{S}=\left(\frac{3\dot{N}_\gamma}{4\pi\alpha_\mathrm{B}n^2_\mathrm{H}}\right)^{1/3}
\end{equation}
is the asymptotically reached Str\"{o}mgren radius. These solutions are valid for 3D space, but similar expressions exist for general dimensional space.

\begin{figure}
\begin{center}
  {\includegraphics[width=\columnwidth,clip=]{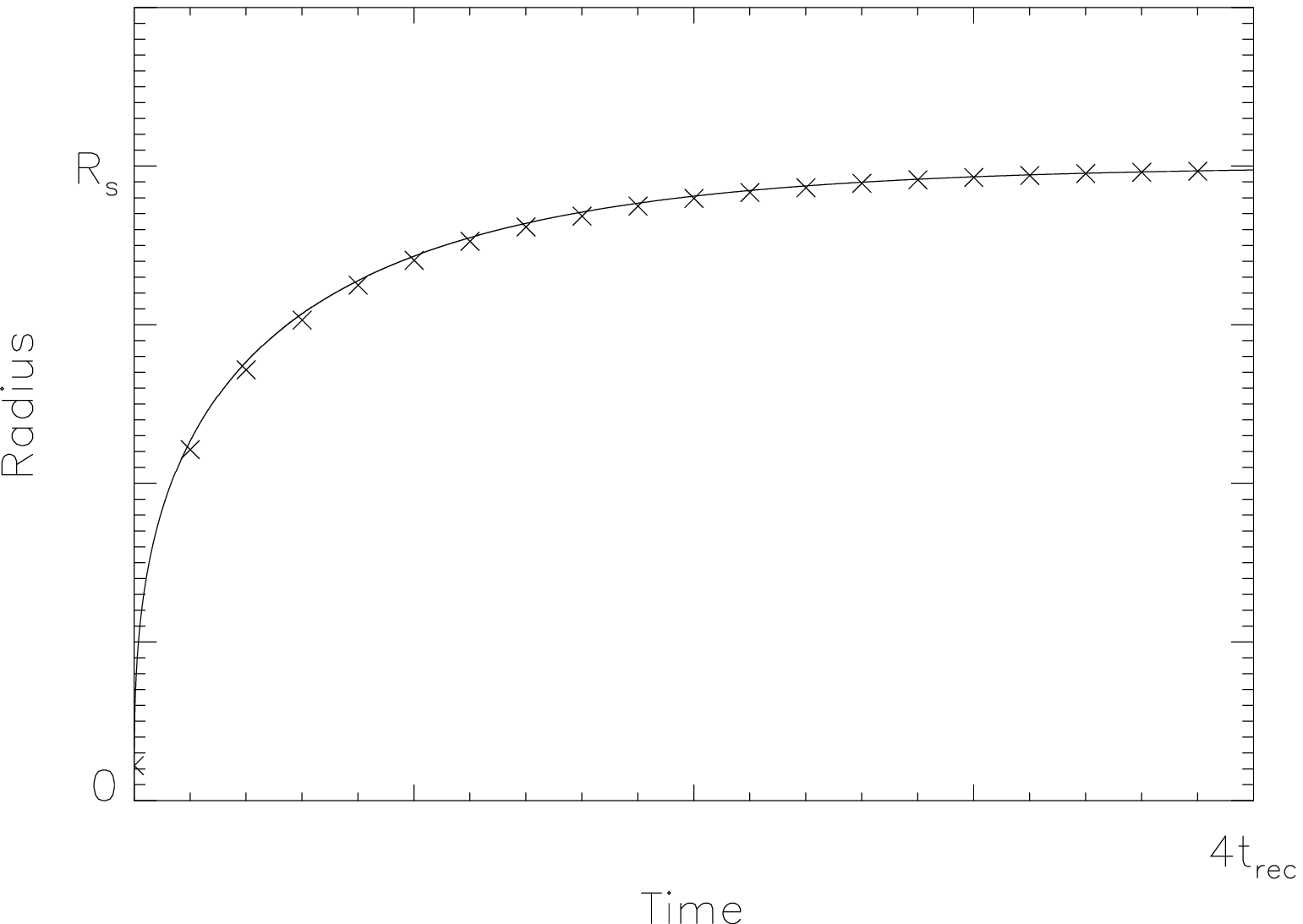}
  \includegraphics[width=\columnwidth,clip=]{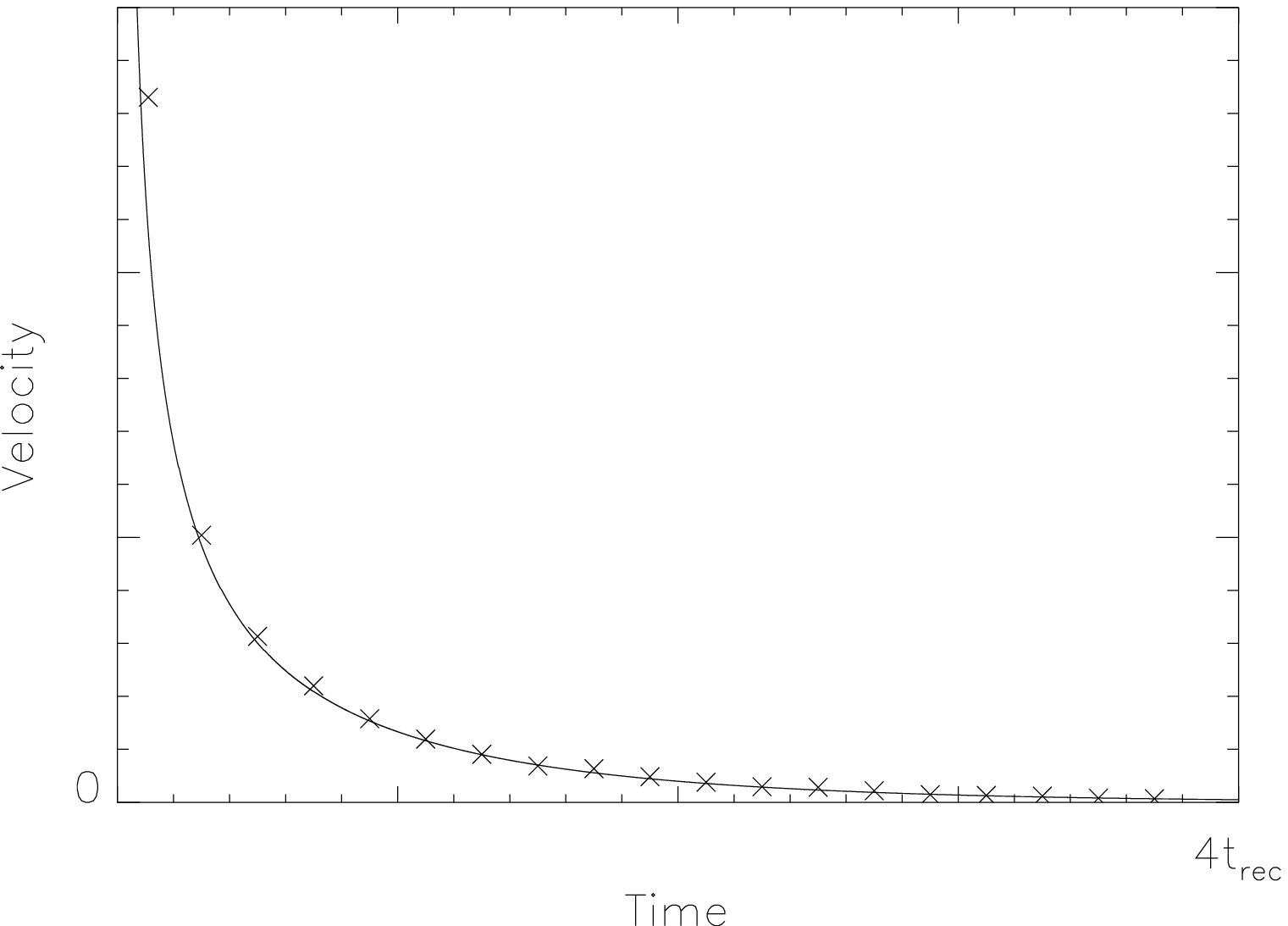}}

  \caption{\label{FigIonFront}Plots of the position (\emph{top}) and velocity (\emph{bottom}) as a function of time for an ionization front in a homogeneous medium. Crosses indicate the data produced by \texttt{SimpleX}, and the solid line represents the analytic solutions Eqs.(\ref{EqIonFrPos}) and (\ref{EqIonFrVel}).
  }
\end{center}
\end{figure}

Taking some typical values, \(\dot{N}_\gamma=5\cdot10^{48}\ \mathrm{s}^{-1}\), \(n_\mathrm{H}=10^{-3}\ \mathrm{cm}^{-3}\), \(\alpha_\mathrm{B}=2.59\cdot10^{-13}\ \mathrm{cm}^3\mathrm{s}^{-1}\) and a chosen simulation time \(T=4t_\mathrm{rec}\), we can run \texttt{SimpleX} with these parameters, and determine what the position and velocity of the ionization front is after each iteration. These results we can immediately compare with the analytical solutions Eqs.(\ref{EqIonFrPos}) and (\ref{EqIonFrVel}), which has been done in Fig.\ \ref{FigIonFront}.

It is straightforward to see that the results of the \texttt{SimpleX} method agree very well with what is analytically expected. A similarly good agreement can be seen, when running the simulation for a medium with an \(r^{-1}\) or an \(r^{-2}\) density distribution, in which \(r\) is the distance to the central source. Analytic solutions are available too for these cases \cite{Mellema}.

Unfortunately, except for these rare cases, there are no analytical solutions available for problems where the medium and the source distribution is more inhomogeneous, and this is precisely the regime in which cosmological radiative transport methods have to perform well. To provide a more quantitative basis on which one can validate the performance of ones code, an cosmological radiative transfer code comparison project was initiated last year, in which \(11\) different codes, of which \texttt{SimpleX} was one, were compared. The comparison consists of several testcases, one of which was described in this subsection, but also several which do not have an analytical solution. The results of this comparison project is described in the forthcoming paper \cite{TSU3}.

\subsection{Large scale simulation}

\begin{figure}
  \begin{center}
  \includegraphics[width=\columnwidth,clip=]{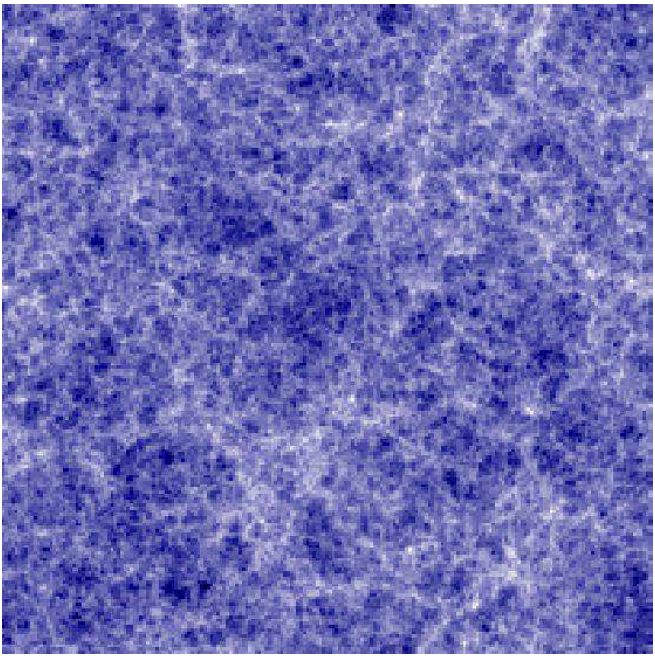}

  \caption{\label{FigDistr}(Color online) A slice through the simulation box, depicting the density distribution of neutral hydrogen on a logarithmic scale. The simulation was done using \texttt{PMFAST} \cite{Merz}. This is a direct representation of the large scale structure of the Universe.}
  
  \end{center}
\end{figure}

As an illustration of what is possible with this method, we give a more realistic example of the reionization of a box which has as a size \(100\ \mathrm{Mpc}\) comoving, in which a certain inhomogeneous matter distribution, extracted from a dark matter simulation with the code \texttt{PMFAST} \cite{Merz}, has to be ionized by a certain number of sources. A slice through the box is depicted in Fig.\ \ref{FigDistr}, in which the logarithm of the density is plotted. One can clearly see the filamentary structure and the high contrast between low and high density regions.

\begin{figure*}
  \begin{center}
  \includegraphics[width=18.0cm,clip=]{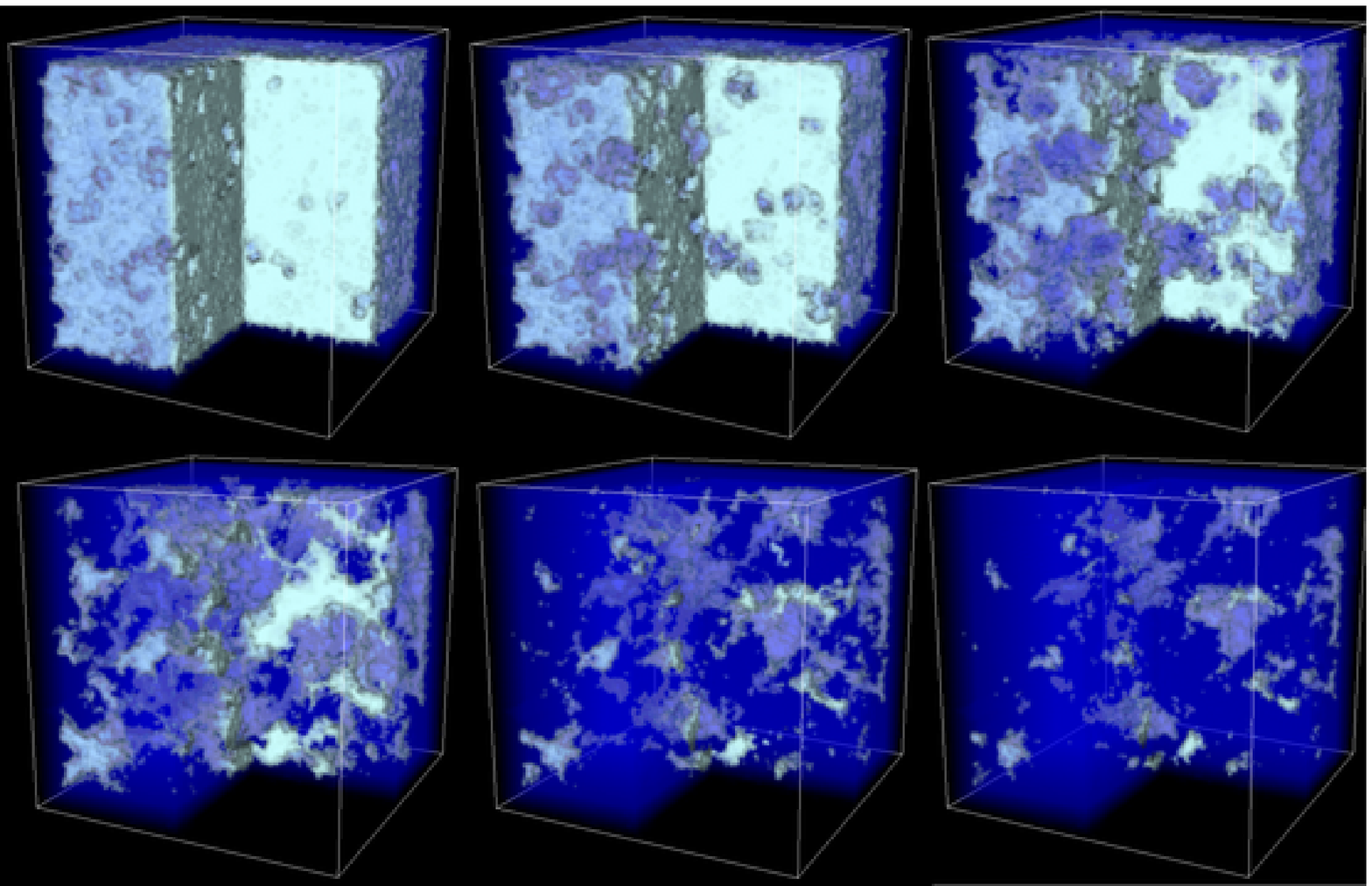}

  \caption{\label{FigCuts}(Color online) A volume rendering of the result of using the \texttt{SimpleX} method to transport the ionizing photons through the medium distribution of the box depicted in Fig.\ \ref{FigDistr}. The results of \(6\) different points in time are plotted, in which the white corresponds to the hydrogen that is still neutral (opaque), and the blue to the already ionized hydrogen (transparent).}
  
  \end{center}
\end{figure*}

We use \(10^6\) points to sample this medium distribution, and via a friends-of-friends halo finding algorithm, we can designate several points as sources of ionizing photons. A total \(2000\) points are marked as sources and we assume that the medium is static, and that there are no radiation hydrodynamical feedback effects.

The result of the \texttt{SimpleX} reionization simulation of this medium distribution is depicted in Fig.\ \ref{FigCuts}, in which a volume rendering of the ionization structure of the medium is plotted for \(6\) different points in time. White is the still neutral and opaque hydrogen, and blue is what is already ionized, and transparent. A cutout has been made to more clearly show the inner structure of the simulated volume.

One can clearly see ionization bubbles, similar to the ones described by Eqs.(\ref{EqIonFrPos}) and (\ref{EqIonFrVel}), being blown around each source, which eventually overlap and fill the whole box. At this point the EOR ends, and, again, the Universe is transparent to ionizing radiation.

\section{Summary and Discussion}

The simulation of the reionization of the large scale structure of the Universe, described in the previous section, runs on a simple desktop computer, finishing within an hour. Other cosmological radiative transfer methods would need weeks, even months of supercomputer time to accomplish the same task. This efficiency is due to the use of the Lagrangian lattice, which puts the usually limited resolution at places where it is needed, whilst at the same time circumventing the severe problems associated with structured grids that introduce unphysical conserved quantities. The resultant method does not scale with the number of sources, which appears to be one of the major bottlenecks of existing linear transport methods. As briefly mentioned in Sect.\ \ref{subsec:CorrelationFunction}, it is possible to include nonlinear terms into our transport process, by discarding the static background medium and letting the random lattice evolve according to the particle density. This paper deals with linear problems only, which is why we refrain from discussing this possible extension here.

In a more general setting, we described in this paper a method which is, in principle, suited to solve linear transport equations in the abstract form of Eq.(\ref{EqAbstractBoltzmann}). Hereto, we assume that we can treat the advection and interaction parts disjointly, or subsequently. The method is new in the sense that it dispenses with the regular, structured grids altogether, and introduces a very physical one.

Focusing on transport in which the only interaction terms involve that between particles and a static medium, we extensively described how to create an adaptive grid, based on a random point process, which locally retains important physical symmetries, namely rotational and translational invariance. Moreover, the resultant lattice can be constructed in such a way that all line lengths correlate linearly with all statistical moments of the path lengths of the particles. As such, the resultant lattice is a direct representation of the free path space of the medium, which characterizes its interaction properties. The exact value of the coupling constant \(\sigma_j\) of each interaction is converted into a global interaction coefficient \(c_j\), which is a quantitative measure of how many mean free paths fit into one lattice line length. After that, we showed that, in the limit of an infinite number of points, the lattice continuously samples all possible free paths and all possible angular directions.

Henceforth, we described how to do the actual transport, by solving for the advection part by moving the particles along the lattice, or graph, and solving for the interaction part, by evaluating Eq.(\ref{EqAbs}) at each point. Redistribution of the surviving particles can be chosen to conserve momentum, and redistribution of scattered particles can be chosen to be isotropic. Of course, it is trivial to incorporate intermediate cases, in which the scattering is anisotropic.

We showed the results of implementing the method as a radiative transfer code \texttt{SimpleX}, with which we evaluated the classical problem of an ionization front expanding in a homogeneous medium towards the asymptotically reached Str\"{o}mgren sphere. The results are in exact agreement with the analytical solutions, which is also the case for \(r^{-1}\) and \(r^{-2}\) medium distributions. The method has been compared to other codes using testcases for which there are no analytical solutions. As an illustration of what the method is capable of, we finished by using our \texttt{SimpleX} method to model the photo-ionization of the large scale structure of the Universe during the Epoch of Reionization.

From an implementation point of view, the method is very easy to use, given that Delaunay and Voronoi tessellation are widely used in many areas of science, including computer visualization, by which fast, robust and open source codes for performing the lattice construction are plentiful. Given the lattice, there is not much more to implement, apart from moving particles along a list of pointers. The expensive finite-differencing, and such, associated with the use of structured grids, was dispensed with together with these lattices. Moreover, because our interaction is implemented as a set of global interaction coefficients, with global interaction rules, each cell can be considered to be a cellular automaton. Thus, our method can be trivially parallelized.

We would like to conclude by noting that although in this paper we focused on linear transport of photon-like particles through a static medium, the method can be used much more generally. The space in which the path lengths are defined does need to be defined as a subset of phase space, but can be much more abstract, and diverse. In every case, the method will consist of constructing a Lagrangian random lattice, as a direct representation of what the transported quantity would encounter, when it is transported through that abstract space.

\begin{acknowledgments}
We are much obliged to Ilian Iliev for making available the dark matter simulation data. JR would like to thank Yuri Levin for carefully reading the manuscript, and Garrelt Mellema, Erik-Jan Rijkhorst, and everyone participating in the comparison project (\texttt{http://www.mpa-garching.mpg.de/tsu3/}) for fruitful discussions.  This work was done with financial support from the Netherlands Organization for Scientific Research (NWO), under grant number 635.000.009.
\end{acknowledgments}

\bibliography{RitzerveldIcke}

\begin{thebibliography}{38}
\expandafter\ifx\csname natexlab\endcsname\relax\def\natexlab#1{#1}\fi
\expandafter\ifx\csname bibnamefont\endcsname\relax
  \def\bibnamefont#1{#1}\fi
\expandafter\ifx\csname bibfnamefont\endcsname\relax
  \def\bibfnamefont#1{#1}\fi
\expandafter\ifx\csname citenamefont\endcsname\relax
  \def\citenamefont#1{#1}\fi
\expandafter\ifx\csname url\endcsname\relax
  \def\url#1{\texttt{#1}}\fi
\expandafter\ifx\csname urlprefix\endcsname\relax\def\urlprefix{URL }\fi
\providecommand{\bibinfo}[2]{#2}
\providecommand{\eprint}[2][]{\url{#2}}

\bibitem[{\citenamefont{Press et~al.}(1992)\citenamefont{Press, Teukolsky,
  Vetterling, and Flannery}}]{NumRecipes}
\bibinfo{author}{\bibfnamefont{W.~H.} \bibnamefont{Press}},
  \bibinfo{author}{\bibfnamefont{S.~A.} \bibnamefont{Teukolsky}},
  \bibinfo{author}{\bibfnamefont{W.~T.} \bibnamefont{Vetterling}},
  \bibnamefont{and} \bibinfo{author}{\bibfnamefont{B.~P.}
  \bibnamefont{Flannery}}, \emph{\bibinfo{title}{Numerical Recipes in C: The
  Art of Scientific Computing}} (\bibinfo{publisher}{Cambridge University
  Press}, \bibinfo{address}{New York, NY, USA}, \bibinfo{year}{1992}).

\bibitem[{\citenamefont{{Richtmyer} and {Morton}}(1967)}]{RichtmyerMorton}
\bibinfo{author}{\bibfnamefont{R.~D.} \bibnamefont{{Richtmyer}}}
  \bibnamefont{and} \bibinfo{author}{\bibfnamefont{K.~W.}
  \bibnamefont{{Morton}}}, \emph{\bibinfo{title}{Difference Methods for Initial
  Value Problems}} (\bibinfo{publisher}{Interscience Publishers},
  \bibinfo{address}{New York}, \bibinfo{year}{1967}).

\bibitem[{\citenamefont{{Van Leer}}(1970)}]{VanLeerThesis}
\bibinfo{author}{\bibfnamefont{B.}~\bibnamefont{{Van Leer}}}, Ph.D. thesis,
  \bibinfo{address}{Leiden} (\bibinfo{year}{1970}).

\bibitem[{\citenamefont{{Icke}}(1988)}]{IckeBubbles}
\bibinfo{author}{\bibfnamefont{V.}~\bibnamefont{{Icke}}},
  \bibinfo{journal}{A\&A} \textbf{\bibinfo{volume}{202}}, \bibinfo{pages}{177}
  (\bibinfo{year}{1988}).

\bibitem[{\citenamefont{{Kunasz} and {Auer}}(1988)}]{Kunasz}
\bibinfo{author}{\bibfnamefont{P.~B.} \bibnamefont{{Kunasz}}} \bibnamefont{and}
  \bibinfo{author}{\bibfnamefont{L.}~\bibnamefont{{Auer}}},
  \bibinfo{journal}{JQSRT} \textbf{\bibinfo{volume}{39}}, \bibinfo{pages}{67}
  (\bibinfo{year}{1988}).

\bibitem[{\citenamefont{{Okabe} et~al.}(1999)\citenamefont{{Okabe}, {Boots},
  {Sugihara}, and {Chiu}}}]{Okabe}
\bibinfo{author}{\bibfnamefont{A.}~\bibnamefont{{Okabe}}},
  \bibinfo{author}{\bibfnamefont{B.}~\bibnamefont{{Boots}}},
  \bibinfo{author}{\bibfnamefont{K.}~\bibnamefont{{Sugihara}}},
  \bibnamefont{and} \bibinfo{author}{\bibfnamefont{S.}~\bibnamefont{{Chiu}}},
  \emph{\bibinfo{title}{Spatial Tessellations, Concepts and Applications of
  Voronoi Diagrams}} (\bibinfo{publisher}{John Wiley \& Sons},
  \bibinfo{year}{1999}), \bibinfo{edition}{2nd} ed.

\bibitem[{\citenamefont{{Christ} et~al.}(1982)\citenamefont{{Christ},
  {Friedberg}, and {Lee}}}]{Christ}
\bibinfo{author}{\bibfnamefont{N.~H.} \bibnamefont{{Christ}}},
  \bibinfo{author}{\bibfnamefont{R.}~\bibnamefont{{Friedberg}}},
  \bibnamefont{and} \bibinfo{author}{\bibfnamefont{T.~D.} \bibnamefont{{Lee}}},
  \bibinfo{journal}{Nuclear Physics B} \textbf{\bibinfo{volume}{202}},
  \bibinfo{pages}{89} (\bibinfo{year}{1982}).

\bibitem[{\citenamefont{{Van Kampen}}(1981)}]{Kampen}
\bibinfo{author}{\bibfnamefont{N.}~\bibnamefont{{Van Kampen}}},
  \emph{\bibinfo{title}{Stochastic Processes in Physics and Chemistry}}
  (\bibinfo{publisher}{North Holland}, \bibinfo{year}{1981}).

\bibitem[{\citenamefont{{Cercignani}}(1988)}]{Cercignani}
\bibinfo{author}{\bibfnamefont{C.}~\bibnamefont{{Cercignani}}},
  \emph{\bibinfo{title}{The Boltzmann Equation and Its Applications}}
  (\bibinfo{publisher}{Springer-Verslag}, \bibinfo{year}{1988}).

\bibitem[{\citenamefont{{Bellomo}}(1995)}]{Bellomo}
\bibinfo{editor}{\bibfnamefont{N.}~\bibnamefont{{Bellomo}}}, ed.,
  \emph{\bibinfo{title}{Lecture Notes on the Mathematical Theory of the
  Boltzmann Equation}} (\bibinfo{publisher}{World Scientific},
  \bibinfo{year}{1995}).

\bibitem[{\citenamefont{{Bird}}(1994)}]{Bird}
\bibinfo{author}{\bibfnamefont{G.}~\bibnamefont{{Bird}}},
  \emph{\bibinfo{title}{Molecular Gas Dynamics and the Direct Simulation of Gas
  Flows}} (\bibinfo{publisher}{Clarendon}, \bibinfo{year}{1994}).

\bibitem[{\citenamefont{{Berger} and {Oliger}}(1984)}]{AMR1}
\bibinfo{author}{\bibfnamefont{M.~J.} \bibnamefont{{Berger}}} \bibnamefont{and}
  \bibinfo{author}{\bibfnamefont{J.}~\bibnamefont{{Oliger}}},
  \bibinfo{journal}{J. Comput. Phys.} \textbf{\bibinfo{volume}{53}},
  \bibinfo{pages}{484} (\bibinfo{year}{1984}).

\bibitem[{\citenamefont{{Berger} and {Colella}}(1989)}]{AMR2}
\bibinfo{author}{\bibfnamefont{M.~J.} \bibnamefont{{Berger}}} \bibnamefont{and}
  \bibinfo{author}{\bibfnamefont{P.}~\bibnamefont{{Colella}}},
  \bibinfo{journal}{J. Comput. Phys.} \textbf{\bibinfo{volume}{82}},
  \bibinfo{pages}{64} (\bibinfo{year}{1989}).

\bibitem[{\citenamefont{{Frisch} et~al.}(1986)\citenamefont{{Frisch},
  {Hasslacher}, and {Pomeau}}}]{FHP}
\bibinfo{author}{\bibfnamefont{U.}~\bibnamefont{{Frisch}}},
  \bibinfo{author}{\bibfnamefont{B.}~\bibnamefont{{Hasslacher}}},
  \bibnamefont{and} \bibinfo{author}{\bibfnamefont{Y.}~\bibnamefont{{Pomeau}}},
  \bibinfo{journal}{\prl} \textbf{\bibinfo{volume}{56}}, \bibinfo{pages}{1505}
  (\bibinfo{year}{1986}).

\bibitem[{\citenamefont{Gladrow}(2000)}]{Wolfe}
\bibinfo{author}{\bibfnamefont{D.~W.} \bibnamefont{Gladrow}},
  \emph{\bibinfo{title}{Lattice Gas and Lattice Boltzmann Methods}}
  (\bibinfo{publisher}{Springer-Verslag, Berlin}, \bibinfo{year}{2000}).

\bibitem[{\citenamefont{Succi}(2001)}]{Succi}
\bibinfo{author}{\bibfnamefont{S.}~\bibnamefont{Succi}},
  \emph{\bibinfo{title}{The Lattice Boltzmann Equation for Fluid Dynamics and
  Beyond}} (\bibinfo{publisher}{Oxford University Press, Oxford},
  \bibinfo{year}{2001}).

\bibitem[{\citenamefont{{Kaku}}(1983)}]{Kaku}
\bibinfo{author}{\bibfnamefont{M.}~\bibnamefont{{Kaku}}},
  \bibinfo{journal}{\prl} \textbf{\bibinfo{volume}{50}}, \bibinfo{pages}{1893}
  (\bibinfo{year}{1983}).

\bibitem[{\citenamefont{{Regge}}(1961)}]{Regge}
\bibinfo{author}{\bibfnamefont{T.}~\bibnamefont{{Regge}}},
  \bibinfo{journal}{Nuovo Cimento A} \textbf{\bibinfo{volume}{19}},
  \bibinfo{pages}{558} (\bibinfo{year}{1961}).

\bibitem[{\citenamefont{Wolfram}(1986)}]{Wolfram}
\bibinfo{author}{\bibfnamefont{S.}~\bibnamefont{Wolfram}}, \bibinfo{journal}{J.
  Stat. Phys.} \textbf{\bibinfo{volume}{45}}, \bibinfo{pages}{471}
  (\bibinfo{year}{1986}).

\bibitem[{\citenamefont{{Ubertini} and {Succi}}(2005)}]{Ubertini}
\bibinfo{author}{\bibfnamefont{S.}~\bibnamefont{{Ubertini}}} \bibnamefont{and}
  \bibinfo{author}{\bibfnamefont{S.}~\bibnamefont{{Succi}}},
  \bibinfo{journal}{Progress in Computational Fluid Dynamics}
  \textbf{\bibinfo{volume}{5}}, \bibinfo{pages}{85 } (\bibinfo{year}{2005}).

\bibitem[{\citenamefont{{Espa{\~n}ol}}(1997)}]{Espanol}
\bibinfo{author}{\bibfnamefont{P.}~\bibnamefont{{Espa{\~n}ol}}},
  \bibinfo{journal}{Europhysics Letters} \textbf{\bibinfo{volume}{39}},
  \bibinfo{pages}{605} (\bibinfo{year}{1997}).

\bibitem[{\citenamefont{{Flekk{\o}y} et~al.}(2000)\citenamefont{{Flekk{\o}y},
  {Coveney}, and {de Fabritiis}}}]{Flekkoy}
\bibinfo{author}{\bibfnamefont{E.~G.} \bibnamefont{{Flekk{\o}y}}},
  \bibinfo{author}{\bibfnamefont{P.~V.} \bibnamefont{{Coveney}}},
  \bibnamefont{and} \bibinfo{author}{\bibfnamefont{G.}~\bibnamefont{{de
  Fabritiis}}}, \bibinfo{journal}{\pre} \textbf{\bibinfo{volume}{62}},
  \bibinfo{pages}{2140} (\bibinfo{year}{2000}).

\bibitem[{\citenamefont{{Delaunay}}(1934)}]{DELAUNAY}
\bibinfo{author}{\bibfnamefont{B.}~\bibnamefont{{Delaunay}}},
  \bibinfo{journal}{Classe des Sciences Math{\'e}matiques et Naturelles}
  \textbf{\bibinfo{volume}{7}}, \bibinfo{pages}{793} (\bibinfo{year}{1934}).

\bibitem[{\citenamefont{{Voronoi}}(1908)}]{VORONOI}
\bibinfo{author}{\bibfnamefont{G.}~\bibnamefont{{Voronoi}}},
  \bibinfo{journal}{Journal f{\"u}r die Reine und Angewandte Mathematik}
  \textbf{\bibinfo{volume}{134}}, \bibinfo{pages}{198} (\bibinfo{year}{1908}).

\bibitem[{\citenamefont{{Garcia} et~al.}(1999)\citenamefont{{Garcia}, {Bell},
  {Crutchfield}, and {Alder}}}]{Garcia}
\bibinfo{author}{\bibfnamefont{A.~L.} \bibnamefont{{Garcia}}},
  \bibinfo{author}{\bibfnamefont{J.~B.} \bibnamefont{{Bell}}},
  \bibinfo{author}{\bibfnamefont{W.~Y.} \bibnamefont{{Crutchfield}}},
  \bibnamefont{and} \bibinfo{author}{\bibfnamefont{B.~J.}
  \bibnamefont{{Alder}}}, \bibinfo{journal}{J. Comput. Phys.}
  \textbf{\bibinfo{volume}{154}}, \bibinfo{pages}{134} (\bibinfo{year}{1999}).

\bibitem[{\citenamefont{{Wu} et~al.}(2002)\citenamefont{{Wu}, {Tseng}, and
  {Kuo}}}]{Wu}
\bibinfo{author}{\bibfnamefont{J.-S.} \bibnamefont{{Wu}}},
  \bibinfo{author}{\bibfnamefont{K.-C.} \bibnamefont{{Tseng}}},
  \bibnamefont{and} \bibinfo{author}{\bibfnamefont{C.-H.} \bibnamefont{{Kuo}}},
  \bibinfo{journal}{Int. J. Numer. Meth. Fluids} \textbf{\bibinfo{volume}{38}},
  \bibinfo{pages}{351} (\bibinfo{year}{2002}).

\bibitem[{\citenamefont{{Benzi} et~al.}(1992)\citenamefont{{Benzi}, {Succi},
  and {Vergassola}}}]{Benzi}
\bibinfo{author}{\bibfnamefont{R.}~\bibnamefont{{Benzi}}},
  \bibinfo{author}{\bibfnamefont{S.}~\bibnamefont{{Succi}}}, \bibnamefont{and}
  \bibinfo{author}{\bibfnamefont{M.}~\bibnamefont{{Vergassola}}},
  \bibinfo{journal}{Physics Reports} \textbf{\bibinfo{volume}{222}},
  \bibinfo{pages}{145} (\bibinfo{year}{1992}).

\bibitem[{\citenamefont{{Karlin} et~al.}(1999)\citenamefont{{Karlin}, {Succi},
  and {Orszag}}}]{Karlin}
\bibinfo{author}{\bibfnamefont{I.~V.} \bibnamefont{{Karlin}}},
  \bibinfo{author}{\bibfnamefont{S.}~\bibnamefont{{Succi}}}, \bibnamefont{and}
  \bibinfo{author}{\bibfnamefont{S.}~\bibnamefont{{Orszag}}},
  \bibinfo{journal}{\prl} \textbf{\bibinfo{volume}{82}}, \bibinfo{pages}{5245}
  (\bibinfo{year}{1999}).

\bibitem[{\citenamefont{{Bradford Barber} et~al.}(1995)\citenamefont{{Bradford
  Barber}, {Dobkin}, and {Huhdanpaa}}}]{QHULL}
\bibinfo{author}{\bibfnamefont{C.}~\bibnamefont{{Bradford Barber}}},
  \bibinfo{author}{\bibfnamefont{D.~P.} \bibnamefont{{Dobkin}}},
  \bibnamefont{and}
  \bibinfo{author}{\bibfnamefont{H.}~\bibnamefont{{Huhdanpaa}}},
  \bibinfo{journal}{ACM Trans. on Mathematical Software}
  \textbf{\bibinfo{volume}{22}}, \bibinfo{pages}{469} (\bibinfo{year}{1995}),
  \urlprefix\url{http://www.qhull.org}.

\bibitem[{\citenamefont{{Loeb} and {Barkana}}(2001)}]{Loeb}
\bibinfo{author}{\bibfnamefont{A.}~\bibnamefont{{Loeb}}} \bibnamefont{and}
  \bibinfo{author}{\bibfnamefont{R.}~\bibnamefont{{Barkana}}},
  \bibinfo{journal}{ARA\&A} \textbf{\bibinfo{volume}{39}}, \bibinfo{pages}{19}
  (\bibinfo{year}{2001}).

\bibitem[{\citenamefont{{Bromm} and {Larson}}(2004)}]{Bromm}
\bibinfo{author}{\bibfnamefont{V.}~\bibnamefont{{Bromm}}} \bibnamefont{and}
  \bibinfo{author}{\bibfnamefont{R.~B.} \bibnamefont{{Larson}}},
  \bibinfo{journal}{ARA\&A} \textbf{\bibinfo{volume}{42}}, \bibinfo{pages}{79}
  (\bibinfo{year}{2004}).

\bibitem[{\citenamefont{{Ciardi} and {Ferrara}}(2005)}]{Ciardi}
\bibinfo{author}{\bibfnamefont{B.}~\bibnamefont{{Ciardi}}} \bibnamefont{and}
  \bibinfo{author}{\bibfnamefont{A.}~\bibnamefont{{Ferrara}}},
  \bibinfo{journal}{Space Science Reviews} \textbf{\bibinfo{volume}{116}},
  \bibinfo{pages}{625} (\bibinfo{year}{2005}).

\bibitem[{\citenamefont{{Ritzerveld}}(2005)}]{Ritzerveld}
\bibinfo{author}{\bibfnamefont{J.}~\bibnamefont{{Ritzerveld}}},
  \bibinfo{journal}{A\&A} \textbf{\bibinfo{volume}{439}}, \bibinfo{pages}{L23}
  (\bibinfo{year}{2005}).

\bibitem[{\citenamefont{{Str{\"o}mgren}}(1939)}]{Stromgren}
\bibinfo{author}{\bibfnamefont{B.}~\bibnamefont{{Str{\"o}mgren}}},
  \bibinfo{journal}{ApJ} \textbf{\bibinfo{volume}{89}}, \bibinfo{pages}{526}
  (\bibinfo{year}{1939}).

\bibitem[{\citenamefont{{Spitzer}}(1978)}]{Spitzer}
\bibinfo{author}{\bibfnamefont{L.}~\bibnamefont{{Spitzer}}},
  \emph{\bibinfo{title}{{Physical processes in the interstellar medium}}}
  (\bibinfo{publisher}{New York Wiley-Interscience, 1978.~333 p.},
  \bibinfo{year}{1978}).

\bibitem[{\citenamefont{{Mellema} et~al.}(2006)\citenamefont{{Mellema},
  {Iliev}, {Alvarez}, and {Shapiro}}}]{Mellema}
\bibinfo{author}{\bibfnamefont{G.}~\bibnamefont{{Mellema}}},
  \bibinfo{author}{\bibfnamefont{I.~T.} \bibnamefont{{Iliev}}},
  \bibinfo{author}{\bibfnamefont{M.~A.} \bibnamefont{{Alvarez}}},
  \bibnamefont{and} \bibinfo{author}{\bibfnamefont{P.~R.}
  \bibnamefont{{Shapiro}}}, \bibinfo{journal}{New Astronomy}
  \textbf{\bibinfo{volume}{11}}, \bibinfo{pages}{374} (\bibinfo{year}{2006}).

\bibitem[{\citenamefont{{Iliev} et~al.}(2006)\citenamefont{{Iliev}, {Ciardi},
  {Alvarez}, {Ferrara}, {Gnedin}, {Maselli}, {Mellema}, {Nakamoto}, {Norman},
  {Razoumov} et~al.}}]{TSU3}
\bibinfo{author}{\bibfnamefont{I.~T.} \bibnamefont{{Iliev}}},
  \bibinfo{author}{\bibfnamefont{B.}~\bibnamefont{{Ciardi}}},
  \bibinfo{author}{\bibfnamefont{M.~A.} \bibnamefont{{Alvarez}}},
  \bibinfo{author}{\bibfnamefont{A.}~\bibnamefont{{Ferrara}}},
  \bibinfo{author}{\bibfnamefont{N.~Y.} \bibnamefont{{Gnedin}}},
  \bibinfo{author}{\bibfnamefont{A.}~\bibnamefont{{Maselli}}},
  \bibinfo{author}{\bibfnamefont{G.}~\bibnamefont{{Mellema}}},
  \bibinfo{author}{\bibfnamefont{T.}~\bibnamefont{{Nakamoto}}},
  \bibinfo{author}{\bibfnamefont{M.~L.} \bibnamefont{{Norman}}},
  \bibinfo{author}{\bibfnamefont{A.~O.} \bibnamefont{{Razoumov}}},
  \bibnamefont{et~al.}, \bibinfo{journal}{ArXiv Astrophysics e-prints}
  (\bibinfo{year}{2006}), \eprint{astro-ph/0603199}.

\bibitem[{\citenamefont{{Merz} et~al.}(2005)\citenamefont{{Merz}, {Pen}, and
  {Trac}}}]{Merz}
\bibinfo{author}{\bibfnamefont{H.}~\bibnamefont{{Merz}}},
  \bibinfo{author}{\bibfnamefont{U.-L.} \bibnamefont{{Pen}}}, \bibnamefont{and}
  \bibinfo{author}{\bibfnamefont{H.}~\bibnamefont{{Trac}}},
  \bibinfo{journal}{New Astronomy} \textbf{\bibinfo{volume}{10}},
  \bibinfo{pages}{393} (\bibinfo{year}{2005}), \eprint{astro-ph/0402443}.

\end{thebibliography}

\begin{appendix}

\section{Conservation of Momentum}
\label{app:ConservationOfMomentum}

\begin{figure}
{\includegraphics[width=\columnwidth]{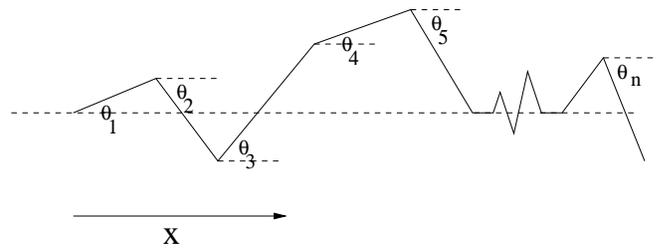}}
\caption{\label{Path_figLong}One possible path of particles performing a walk
of \(n\) steps on the Delaunay graph. The \(i\)-th step is parametrized by an angle \(\theta _i\),
with respect to the original direction \(\mathbf{x}\).}
\end{figure}
The symmetry of each typical Voronoi cell \cite{Okabe} ensures that momentum is conserved on average, given the ballistic transport recipe in Sect.\ \ref{subsec:BallisticTransport}. What is more interesting, though, from an implementation point of view, is the width of the distribution around the original direction. That is, how wide would an otherwise infinitely thin laser beam become as function of the distance to the source. For an exact mathematical analysis hereof for the Poissonian random lattice, we proceed as follows. 

An example of a path of a particle performing a walk in two dimensions is given in Fig.\ \ref{Path_figLong}. The following analysis, however, will be valid in \( d \)-dimensional space.
Because of cylindrical symmetry around the original direction  \( \mathbf{x} \) (the distribution function of the deflection angle is symmetric), we can parametrize the
\( i \)-th step of the particle's walk by only one angle \( \theta _{i} \), which is the angle between the \( i \)-th Delaunay edge and the original direction \(\mathbf{x}\).
Thus, the expectation value of the total displacement \( \mathbf{R}_{n}=\mathbf{r}_{1}+...+\mathbf{r}_{n} \) is
\begin{eqnarray}
  \label{vec_average_long}
  \left\langle \mathbf{R}_{n}\right\rangle & = & \left\langle \mathbf{r}_{1}\right\rangle +...+\left\langle   \mathbf{r}_{n}\right\rangle \nonumber \\
  & = & {n}\left<L\right>\left\langle \cos \theta \right\rangle \frac{\mathbf{x}}{\left| \mathbf{x}\right| } \nonumber \\
  & = & {n}\left<L\right>\chi \frac{\mathbf{x}}{\left| \mathbf{x}\right| },
\end{eqnarray}
in which \( \left<L\right> \) is the average Delaunay line length, defined in Eq.(\ref{EqExpectedDelaunayLineLengthK}), and
\begin{equation}
\label{chi_defined}
\chi = \int ^{\pi} _{-\pi} h(\theta)\cos\theta d\theta.
\end{equation}
Here, we have used \( h(\theta ) \) as a certain symmetric function, which characterises the probability distribution of the angle \(\theta\) and which, in most cases, cannot be evaluated analytically. Several Monte Carlo experiments for this angle have been done \cite{Okabe}.

The second-order expectation value can be evaluated as follows:
\begin{eqnarray}
\label{sec_exp_long}
\left\langle \mathbf{R}^{2}_{n}\right\rangle  & = & \left\langle \mathbf{r}^{2}_{1}\right\rangle +\left\langle \mathbf{r}_{1}\cdot
\mathbf{r}_{2}\right\rangle +...+\left\langle \mathbf{r}^{2}_{n}\right\rangle \nonumber \\
 & = &\left<L^2\right>\left[{n} + {n}({n-1})\left\langle \cos ( \theta _{i} + \theta _{j} ) \right\rangle\right],
\end{eqnarray}
in which we may choose \( i \) and \( j \) randomly from the set \( \left\{ 1,...,n \right\} \), as long as
\( i \not = j \), because the distribution function \( h(\theta ) \) has the same form for
each angle \( \theta _i \). Using the cosine addition formula, we can reduce Eq.(\ref{sec_exp_long}) to
\begin{equation}
\label{sec_explong_final}
\left\langle \mathbf{R}^{2}_{n}\right\rangle = \left({n} + {n}({n-1})\chi^2\right)\left<L^2\right>.
\end{equation}
Thus, the variance of the displacement is
\begin{equation}
\label{variance_long}
\sigma ^2 _{\mathbf{R} _{n}} = {n} \left<L^2\right>(1-\chi ^2).
\end{equation}
When \( h( \theta ) \propto \delta(\theta) \), then \( \chi = 1 \), by which \( \left\langle \mathbf{R} _{n} \right\rangle =\left<L\right> {n}\mathbf{x}/\left| \mathbf{x}\right| \) 
and \( \sigma ^2 _{\mathbf{R} _{n}} = 0 \) as should be expected.
The exact form of a distribution function like \( h(\theta) \)
can probably not be evaluated, even for this well-studied Poisson case, but we can use a step function as an approximation. 
Thus, given that in \(\mathbb{R}^2\) the average number of Delaunay lines meeting at a grid point is \(6\), we use as a step function
\( h(\theta) = 3 / \pi \) on the domain \( \theta \in [-\pi/6,\pi/6] \).
This results in \( \chi = 3/\pi \), by which 
\begin{equation}
\label{exp_long_step}
\left\langle \mathbf{R}_{n} \right\rangle = \frac{3n\left<L\right>}{\pi} \frac{\mathbf{x}}{\left| \mathbf{x}\right| },
\end{equation}
which is very close (difference of less than \(5\%\))
to the distance along a straight line, which would be \(n\left<L\right>\). We can always, of course, 
rescale the lengths so as to make sure that the distance traversed equals the exact physical one.

More importantly, the variance in the displacement, in this case, is
\begin{equation}
\label{variance_long_step}
\sigma ^2 _{\mathbf{R} _{n}} = \frac{\pi^2 - 9}{\pi^2} \left<L^2\right>{n}.
\end{equation}
We know that the results of using a step-function as distribution function gives upper bounds
on the values of Eqs.(\ref{vec_average_long}) and (\ref{variance_long}), because, similar to the deflection angle distribution function,  the actual
distribution function would peak around \( \theta = 0 \) and would decrease as \( \left| \theta \right| \) 
increases, so we expect the actual value of \( \sigma ^2 _{\mathbf{R} _{n}} \) to be smaller.
Thus, we can simulate a straight line trajectory with this method, because \( \mathbf{R} _{n}\propto{n}\mathbf{x}\), but with a standard deviation that increases with \( \sqrt{n} \).

A crucial aspect is the behavior of the standard deviation, when the number of
grid points \( N \) increases. Let us therefore examine a line segment in the simulation
domain of length \( \mathcal{L} \) (\(\leq \sqrt{d}\), if we have a \([0.0:1.0]^d\) domain). 
Because the point distribution is homogeneous, we can conclude 
that the number of steps to cover the line is
\begin{equation}
\label{line_steps}
{n}=\xi {N}^{1/d},
\end{equation}
in which \( \xi \leq \frac{\pi}{3}\zeta(1,2)\sqrt{d} \), which can be found by using the upper bound Eq.(\ref{exp_long_step}) and
the Eq.(\ref{EqExpectedDelaunayLineLengthK}) for the length \(\left<L\right>\) of a Delaunay line.
If we combine Eq.(\ref{variance_long_step}) with Eq.(\ref{line_steps}), again using Eq.(\ref{EqExpectedDelaunayLineLengthK}), we obtain
\begin{equation}
\label{sigma_long_number}
\sigma \propto \left<L\right> \sqrt{n} \propto {N}^{-1/2d}.
\end{equation}
Thus, we can conclude that the amount of widening of the beam will go to zero, when we increase the amount of grid points \(N\).

Even if we do not have a large amount of points to suppress the widening of the beam,
we have another effect which compensates for the widening. Namely, at each intersection
the number of particles are split up into \( d \) parts. This means that the particle number at points
farther away from the straight line trajectory is much less than at points close by, simply
because of the fact that more paths cross each other at points close to the line.

\end{appendix}

\end{document}